\documentclass[twocolumn,english,aps,prb,floats]{revtex4-2}
\usepackage[T1]{fontenc}
\usepackage[latin9]{inputenc}
\usepackage{babel}
\usepackage{textcomp}
\usepackage{amsmath}
\usepackage{amssymb}
\usepackage{graphicx}
\usepackage{wasysym}
\usepackage{esint}
\usepackage[unicode=true,
 bookmarks=false,
 breaklinks=false,pdfborder={0 0 1},backref=section,colorlinks=false]
 {hyperref}

\makeatletter

\newcommand{\lyxmathsym}[1]{\ifmmode\begingroup\def\b@ld{bold}
  \text{\ifx\math@version\b@ld\bfseries\fi#1}\endgroup\else#1\fi}

\usepackage{babel}

\@ifundefined{textcolor}{}{%
 \definecolor{BLACK}{gray}{0}
 \definecolor{WHITE}{gray}{1}
 \definecolor{RED}{rgb}{1,0,0}
 \definecolor{GREEN}{rgb}{0,1,0}
 \definecolor{BLUE}{rgb}{0,0,1}
 \definecolor{CYAN}{cmyk}{1,0,0,0}
 \definecolor{MAGENTA}{cmyk}{0,1,0,0}
 \definecolor{YELLOW}{cmyk}{0,0,1,0}
}

\usepackage{ifpdf}\usepackage{bm}

\makeatother

\begin{document}
\title{Polyhexatic and Polycrystalline States of Skyrmion Lattices}
\author{Dmitry A. Garanin and Eugene M. Chudnovsky}
\affiliation{Physics Department, Herbert H. Lehman College and Graduate School,
The City University of New York, 250 Bedford Park Boulevard West,
Bronx, New York 10468-1589, USA }
\date{\today}
\begin{abstract}
We report Monte Carlo studies of lattices of up to $10^{5}$ skyrmions
treated as particles with negative core energy and repulsive interaction
obtained from a microscopic spin model. Temperature dependence of
translational and orientational correlations has been investigated
for different experimental protocols and initial conditions. Cooling
the skyrmion liquid from a fully disordered high-temperature state
results in the formation of a skyrmion polycrystal. A perfect skyrmion
lattice prepared at $T=0$, on raising temperature undergoes first
order melting transition into a polyhexatic state that consists of
large orientationally ordered domains of fluctuating shape. On further
increasing temperature these domains decrease in size, leading to
a fully disordered liquid of skyrmions. 
\end{abstract}
\maketitle

\section{Introduction}

Skyrmions came to material science from nuclear physics where they
were introduced as solutions of the nonlinear $\sigma$-model that
can describe nucleons, deuterons, $\alpha$-particles and more complex
atomic nuclei \citep{SkyrmePRC58,Polyakov-book,Manton-book,D1,D2}.
They possessed a topological charge, $Q=\pm1,\pm2$, etc., that was
identified with the baryon number. Mathematical equivalence of the
nonlinear $\sigma$-model to the exchange model of ferro- and antiferromagnets
injected skyrmions into condensed matter physics. In magnetic films,
they are defects of the magnetic order. Their topological charge $Q$
corresponds to a homotopy class of the mapping of a three-component
fixed-length magnetization field (or a Néel vector) onto a 2D plane
of the film.

Skyrmions in magnetic systems have been actively studied in recent
years due to interesting physics they entail and beautiful images
they provide, but also due to their potential for developing topologically
protected nanoscale information carriers that can be manipulated by
electric currents \citep{Nagaosa2013,Zhang2015,Klaui2016,Hoffmann-PhysRep2017,Fert-Nature2017}.
Much larger cylindrical domains surrounded by thin domain walls --
magnetic bubbles, with similar topological properties, were intensively
investigated by magneto-optical methods in 1970s \citep{MS-bubbles,ODell}.
On the contrary, magnetic skyrmions can be small compared to the domain
wall thickness, making them objects of nanoscience that are conceptually
similar to the topological objects studied in nuclear physics \citep{BelPolJETP75,Lectures}.

The work on magnetic skyrmions focused on their stability and phase
diagrams separating skyrmion states from other magnetic structures.
In a pure exchange model of the magnetic order in a 2D solid, skyrmions
collapse \citep{CCG-PRB2012} due to violation of the scale invariance
by the atomic lattice. Macroscopic arrays of magnetic bubbles observed
in the past were in effect domain structures stabilized by the perpendicular
magnetic anisotropy (PMA), dipole-dipole interaction (DDI), and the
external magnetic field \citep{MS-bubbles,ODell,Ezawa-PRL2010,Makhfudz-PRL2012}.
Since skyrmions are much smaller, their stability requires additional
interactions. At sufficiently low temperature, individual skyrmions
can be stabilized by Dzyaloshinskii-Moriya interaction (DMI) \citep{Bogdanov1989,Bogdanov94,Bogdanov-Nature2006,Heinze-Nature2011,Boulle-NatNano2016,Leonov-NJP2016}
in materials lacking inversion symmetry. Stability of the skyrmions
can also be provided by frustrated exchange interactions \citep{Leonov-NatCom2015,Zhang-NatCom2017},
magnetic anisotropy \citep{IvanovPRB06,Lin-PRB2016}, disorder \citep{CG-NJP2018},
and geometrical confinement \citep{Moutafis-PRB2009}. Thermal fluctuations
destroy nanoscale skyrmions with the rate depending on temperature
and the energy barrier separating stable and unstable skyrmion states
\citep{bessarab18,stosic17,desplat18,hagemeister15,rohart16,rozsa16,siemens16,Malottki-PRB2019,Amel-JAP2019,Hoffmann-PRL2020}.

Two-dimensional lattices of magnetic vortices in systems with DMI
and uniaxial magnetic anisotropy were initially investigated by Bogdanov
and Hubert \citep{Bogdanov94} within continuous micromagnetic theory.
By comparing energies of the periodic arrangement of vortices with
energies of laminar domains, they obtained the magnetic phase diagram.
Their findings received experimental confirmation from imaging of
magnetic phases in FeCoSi films with the help of the real-space Lorentz
transmission electron microscopy \citep{Yu2010}. Transitions between
uniformly magnetized states, laminar domains, and skyrmion crystals
on temperature and the magnetic field were observed and confirmed
by many studies that followed, see, e.g., Refs. \citep{GCZZ-MMM2021,Dohi-ARCMP2022},
and references therein.

Observation of hexagonal skyrmion lattices has provided a new area
for the studies of melting of 2D solids. This problem was first addressed
in the 1970s in seminal works of Halperin and Nelson \citep{HN-PRL1978,NH-PRB1979},
and Young \citep{Young-PRB1979} (see also Ref. \citep{Strandburg}
for review and references therein). They showed that in accordance
with the Kosterlitz-Thouless (KT) theory \citep{KT}, a 2D solid melts
via unbinding of defects of the order parameter, in this case dislocations.
However, unlike the melting transition in continuous 2D models of,
e.g., a 2D ferromagnet or a superconductor, a 2D crystal was found
to melt in a two-step manner. First it melts into an intermediate
hexatic phase with exponential decay of translational correlations
but algebraic decay of orientational correlations. Then, on further
increasing the temperature, it melts in a true liquid state with exponential
decay of all correlations.

The KTHNY has been confirmed by simulations of 2D atomic lattices
\citep{DC-PRB1995,Li-PRE2019} and in experiments on lattices formed
by colloidal particles \citep{Murray-PRB1990,Keim}. However, the
full clarity about the phase diagram of a 2D solid has not been achieved.
First-order melting transition has been observed in vortex lattices
of high-temperature superconductors \citep{Zeldov-1995}. Early molecular-dynamics
studies \citep{Broughton1982} elucidated the importance of the competition
between long-wave fluctuations that contribute to the KTHNY two-step
melting and short-wave phonons that together with lattice defects
drive a conventional first-order melting. Which one prevails over
the other depends on the interaction potential \citep{Broughton1982,Tsiok2022,Kapfer2015}
and the symmetry of the lattice \citep{Janke1988,Dietel2006}.

Skyrmion lattices represent the most recent tool for testing the theory
of 2D melting. Using large-scale Monte Carlo simulations of a system
of classical spins on the lattice, Nishikawa et al. \citep{Nishikawa-PRB2019}
observed the direct melting of the skyrmion lattice into a 2D liquid
with short-range correlations, with no intermediate hexatic phase.
However, Huang et al. \citep{Huang-Nat2020}, using cryo-Lorentz transmission
electron microscopy of a Cu$_{2}$OSeO$_{3}$ nano-slab, reported
a two-step melting transition of the skyrmion lattice, with hexatic
phase sandwiched in the phase diagram between a 2D solid and a 2D
liquid. Similar conclusion was reached by Baláž et al. \citep{Balaz-PRB2021}
by simulating skyrmion lattices in a GaV$_{4}$S$_{8}$ spinel. Metastability
and hysteresis in the dynamics of skyrmion lattices, caused by slow
relaxation, has been emphasized in recent works of Zazvorka et al.
\citep{Zazvorka2020} and McCray et el. \citep{McCray2022}.

While modern numerical simulations of 2D lattices study systems of
size up to $10^{6}$ particles \citep{Kapfer2015}, simulations of
skyrmion lattices in terms of spins on a lattice are more demanding
as one skyrmion comprizes many spins. The largest system studied in
Ref. \citep{Nishikawa-PRB2019} has $10^{6}$ lattice spins but only
$128\times128=16384$ small skyrmions. In this paper, a simplified
treatment of skyrmions as particles is proposed using the recently
established repulsive skyrmion-skyrmion interaction \citep{CGC-JPCM2020}.
This has allowed us to simulate systems consisting of a much larger
number of skyrmions. Unlike most of the work done on particles with
power-law interaction potentials, the repulsion between skyrmions
decreases exponentially with the distance.

In our Monte Carlo simulations, we study both melting of a 2D skyrmion
single crystal on increasing temperature and solidification of the
skyrmion liquid on lowering temperature. In the latter case, unless
a specific crystal-growth condition is implemented, a sufficiently
large system of particles always breaks into randomly oriented crystallites
with no global orientational order. This must apply to the systems
of skyrmions as well, regardless of whether they emerge in a 2D film
on lowering temperature or from laminar magnetic domains on changing
the magnetic field.

We show that, on increasing the temperature, a single-crystalline
or polycrystalline lattice melts into the skyrmion liquid via the
intermediate state that we call \textit{polyhexatic state}. This state
consists of grains of lattice with the same orientations of hexagons.
We show that, on raising the temperature, a monocrystalline or a polycrystalline
skyrmion solid melts into a skyrmion liquid via an intermediate state
that we call the polyhexatic state. It consists of domains with the
same orientations of hexagons inside each domain but different orientations
of hexagons from one domain to the other. Unlike frozen boundaries
between crystallites observed at low temperature, the boundaries between
domains in a polyhexatic state fluctuate. On a larger time scale,
grains appear and disappear. On further increasing the temperature
above the melting transition, the domains of orientational order gradually
decrease in size, leading to a fully disordered skyrmion liquid. We
observe a single first-order transition from a skyrmion solid to a
polyhexatic state.

The paper is organized as follows. The magnetic model on a lattice,
including skyrmions, repulsion between them, and construction of skyrmion
lattices for different system shapes is introduced in Section \ref{Sec-Skyrmion}.
In Sec. \ref{Sec-Physical-quantities} physical quantities such as
orientational and translational order parameters and correlation functions
(CF) are discussed. The discussion of lattice defects and their numerical
detection is given at the end of this section. The application of
Monte Carlo method to the skyrmion lattice at nonzero temperatures
is described in Sec. \ref{Sec-Numerical-method}. Our numerical findings
are presented in Section \ref{Sec-Numerical_results}. Their relevance
to the theory of 2D melting and experiments on skyrmion lattices are
discussed in Section \ref{Sec-Conclusion}.

\section{Single skyrmion and skyrmion lattice}

\label{Sec-Skyrmion}

\subsection{The model and properties of a single skyrmion}

We start with a model of ferromagnetically coupled three-component
classical spin vectors ${\bf s}_{i}$ on a square lattice having the
energy: 
\begin{eqnarray}
\mathcal{H} & = & -\frac{1}{2}\sum_{ij}J_{ij}\mathbf{s}_{i}\mathbf{s}_{j}-H\sum_{i}s_{iz}\nonumber \\
 &  & -A\sum_{i}\left[(\mathbf{s}_{i}\times\mathbf{s}_{i+\delta_{x}})_{y}-(\mathbf{s}_{i}\times\mathbf{s}_{i+\delta_{y}})_{x}\right].\label{Hamiltonian}
\end{eqnarray}
Here $\left|{\bf s}_{i}\right|=1$ and the exchange coupling $J>0$,
incorporating the actual length of the spin, is limited to the nearest
neighbors. It favors ferromagnetic ordering that we chose in the negative
(downward) \textit{z}-direction at infinity (or at the boundary of
a finite-area film). The stabilizing field $H$ in the second (Zeeman)
term is applied in the same negative \textit{z}-direction, $H<0$,
to support this configuration. The third term in Eq.\ (\ref{Hamiltonian})
is Dzyaloshinskii-Moriya interaction of the Néel-type, where $\delta_{x}$
and $\delta_{y}$ refer to the next nearest lattice site in the positive
$x$ or $y$ direction. 

Spin-field configurations in 2D belong to homotopy classes characterized
by the topological charge that in the continuous approximation is
defined as
\begin{equation}
Q=\int\frac{dxdy}{4\pi}\:{\bf s}\cdot\frac{\partial{\bf s}}{\partial x}\times\frac{\partial{\bf s}}{\partial y}\label{Q}
\end{equation}
and takes quantized values $Q=0,\pm1,\pm2,...$. Exact analytical
solutions for spin states in the pure-exchange model in the continuous
approximation were found by Belavin and Polyakov\citep{BelPolJETP75}
(see a simplified derivation in Ref. \citep{CGC-JPCM2020}). These
states have the positive energy $\Delta E_{\mathrm{BP}}=4\pi J|Q|$
with respect to the uniform magnetic state. Conservation of the topological
charge prevents an easy conversion of these states into the uniform
state. The most important of these states are skyrmions and antiskyrmions
that in the configuration with downward spins at infinity have $Q=\pm1$,
respectively. The in-plane components of the spin field in the skyrmion
have the form
\begin{equation}
\left\{ \begin{array}{c}
s_{x}(r,\phi)\\
s_{y}(r,\phi)
\end{array}\right\} =\sqrt{1-s_{z}^{2}(r)}\left\{ \begin{array}{c}
\cos\left(\phi+\gamma\right)\\
\sin\left(\phi+\gamma\right)
\end{array}\right\} ,\label{Skyrmion}
\end{equation}
while $s_{z}(r)$ changes between 1 in the skyrmion's center, $r=0$
and $-1$ at $r=\infty$. Here $\gamma$ is the chirality: $\gamma=0$
for the outward Néel skyrmion, $\gamma=\pi$ for the inward Néel skyrmion
and $\gamma=\pm\pi/2$ for Bloch skyrmions. For the pure-exchange
model \citep{BelPolJETP75}
\begin{equation}
s_{z}(r)=\frac{\lambda^{2}-r^{2}}{\lambda^{2}+r^{2}},
\end{equation}
where $\lambda$ is the skyrmion size. Note that the energy $\Delta E_{\mathrm{BP}}$
of the Belavin-Polyakov (BP) skyrmion does not depend on $\lambda$
and $\gamma$. Discreteness of the lattice breaks this invariance
adding a term of the order $-\left(a/\lambda\right)^{2}$, where $a$
is the lattice spacing, to the skyrmion's energy that leads to the
skyrmion collapse \citep{CCG-PRB2012}.

\begin{figure}

\begin{centering}
\includegraphics[width=9cm]{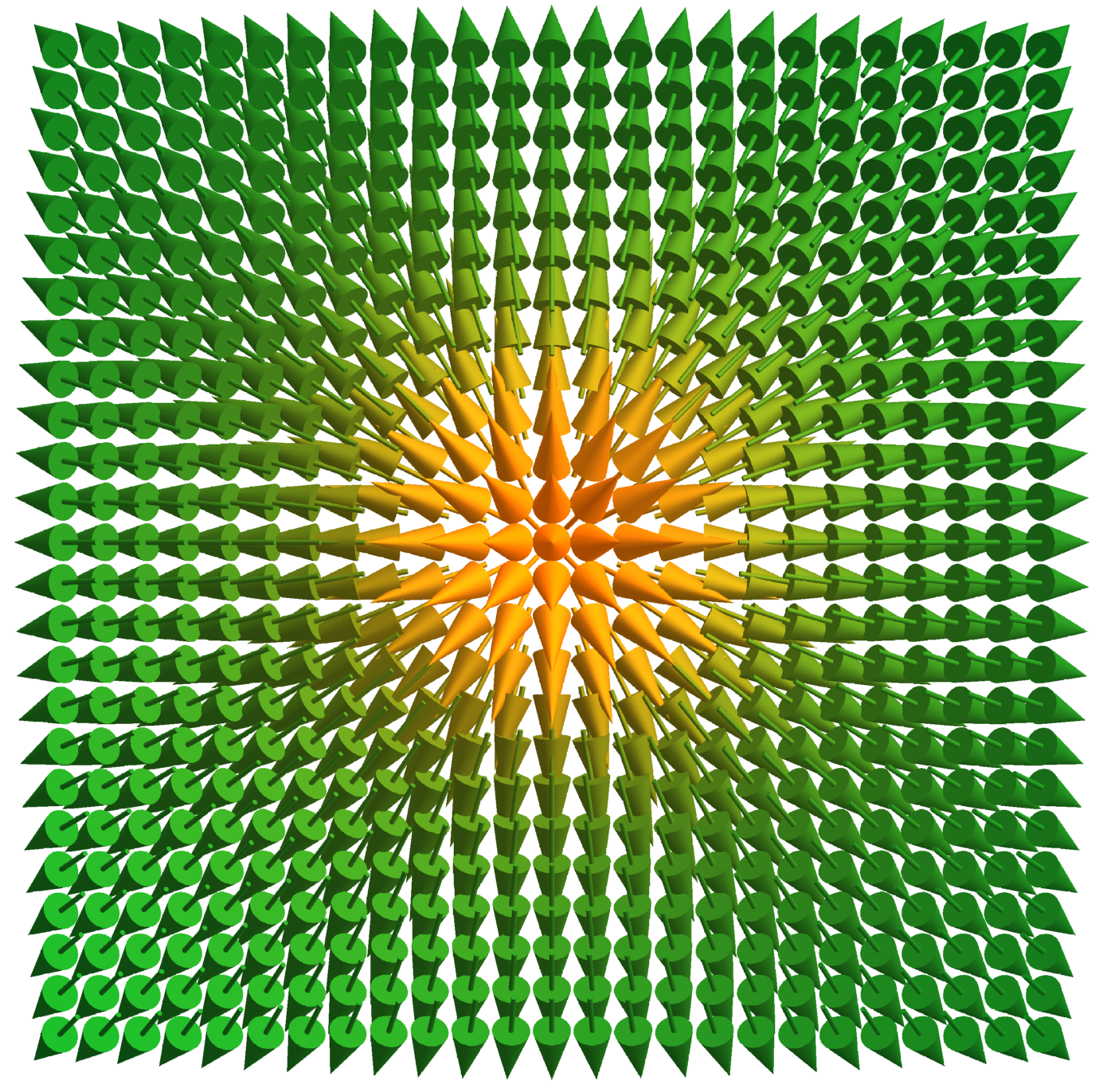}\caption{An outward Néel skyrmion, view from above (in the negative $z$-direction).}
\par\end{centering}
\label{Fig-N-skyrmion}
\end{figure}

\begin{figure}
\begin{centering}
\includegraphics[width=9cm]{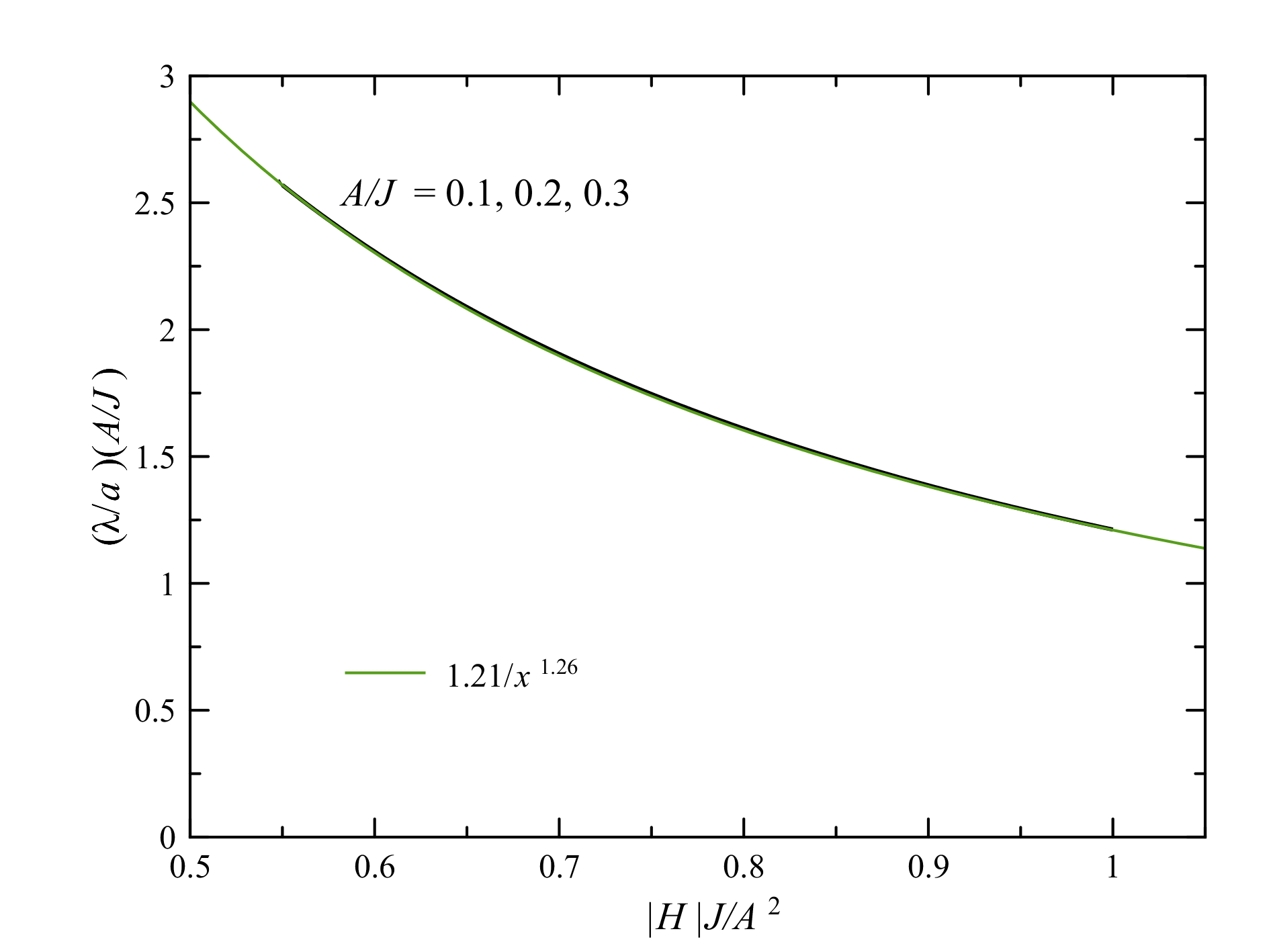}
\par\end{centering}
\caption{The scaled plot of the skyrmion size $\lambda$ vs. $H$ for different
values of $A$.}

\label{Fig-=0003BB_vs_H_scaled_scaled}
\end{figure}

\begin{figure}
\begin{centering}
\includegraphics[width=9cm]{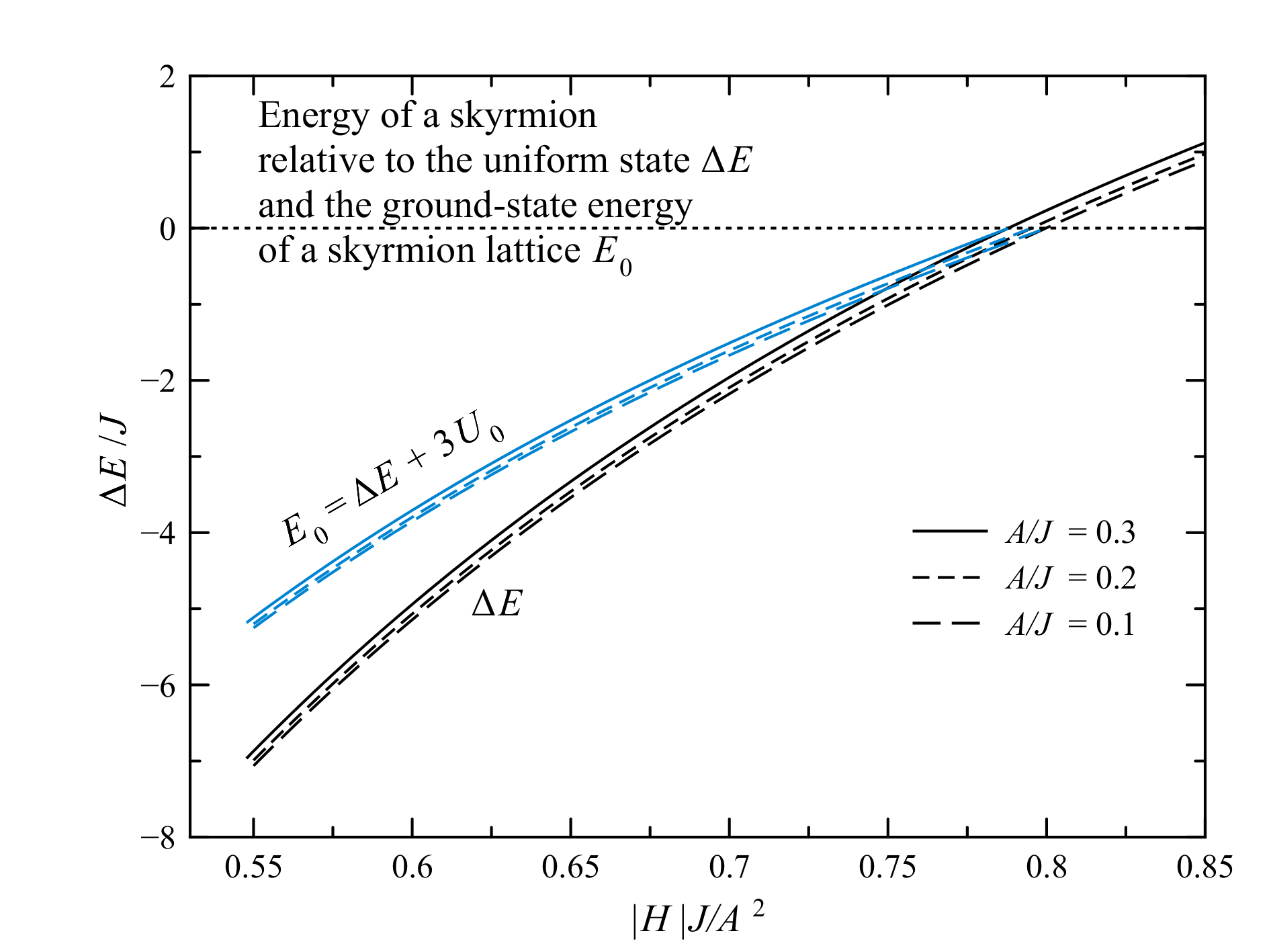}
\par\end{centering}
\caption{The energy of a single skyrmion $\Delta E$ with respect to the uniform
state and the ground-state energy of an equilibrium skyrmion lattice
per skyrmion $E_{0}$ vs $H$.}

\label{Fig-Delta_E_vs_H_scaled}
\end{figure}
In the presence of the DMI, only skyrmions can exist as nontrivial
topological states. The Néel-type DMI above favors outward skyrmions
for $A>0$ (see Fig. \ref{Fig-N-skyrmion}), and inward skyrmions
for $A<0$. There is another type of DMI that favors Bloch skyrmions.
Most of the results are the same for both types of DMI. There is no
analytical solution for the skyrmion's profile $s_{z}(r)$ for the
model with the DMI. In the numerical energy minimization on a lattice,
the skyrmion size $\lambda$ can be defined using the formula \citep{CCG-PRB2012}:
\begin{equation}
\lambda_{n}^{2}=\frac{n-1}{2^{n}\pi}a^{2}\sum_{i}(s_{i,z}+1)^{n},\label{Lambda}
\end{equation}
that yields $\lambda_{n}=\lambda$ for the BP skyrmions for any $n$.
For the computations, we took $n=4$ to give more weight to the skyrmion's
core. 

To find the single-skyrmion spin configuration at $T=0$ numerically,
one can start with a BP skyrmion or any topologically similar structure
and perform energy minimization. Our numerical method combines sequential
rotations of spins ${\bf s}_{i}$ towards the direction of the local
effective field, ${\bf H}_{{\rm eff},i}=-\partial{\cal H}/\partial{\bf s}_{i}$,
with the probability $\alpha$, and the energy-conserving spin flips
(so-called \textit{overrelaxation}), ${\bf s}_{i}\to2({\bf s}_{i}\cdot{\bf H}_{{\rm eff},i}){\bf H}_{{\rm eff},i}/H_{{\rm eff},i}^{2}-{\bf s}_{i}$,
with the probability $1-\alpha$. The parameter $\alpha$ plays the
role of the effective relaxation constant. We mainly use the value
$\alpha=0.03$ that provides the overall fastest convergence. We used
the system dimensions $100\times100$ for $A/J=0.3,$0.2 and $200\times200$
for $A/J=0.1$. The scaled plot of the skyrmion size $\lambda$ vs
$H$ for different values of $A$ is shown in Fig. \ref{Fig-=0003BB_vs_H_scaled_scaled}.
The skyrmion solution exists in the field interval $H_{s}\leq|H|\leq H_{c}$,
where $H_{s}\simeq0.55A^{2}/J$ is the ``strip-out'' field below
which the skyrmion becomes unstable and converts into a laminar domain
structure and $H_{c}\simeq0.97A^{3/2}/J^{1/2}$ \citep{derchugar2018}
is the skyrmion-collapse field. The ratio of the field boundary values
is $H_{c}/H_{s}\simeq1.76\left(J/A\right)^{1/2}$ that is greater
than one for realistic parameters' values.

The energy of a single skyrmion with respect to the uniform state
becomes negative for $A$ large enough and $|H|$ small enough that
makes the skyrmion thermodynamically stable, see black lines in Fig.
\ref{Fig-Delta_E_vs_H_scaled}.

\subsection{Skyrmion-skyrmion repulsion and the equilibrium skyrmion lattice}

\label{subsec:Skyrmion-skyrmion repulsion}

\begin{figure}
\begin{centering}
\includegraphics[width=9cm]{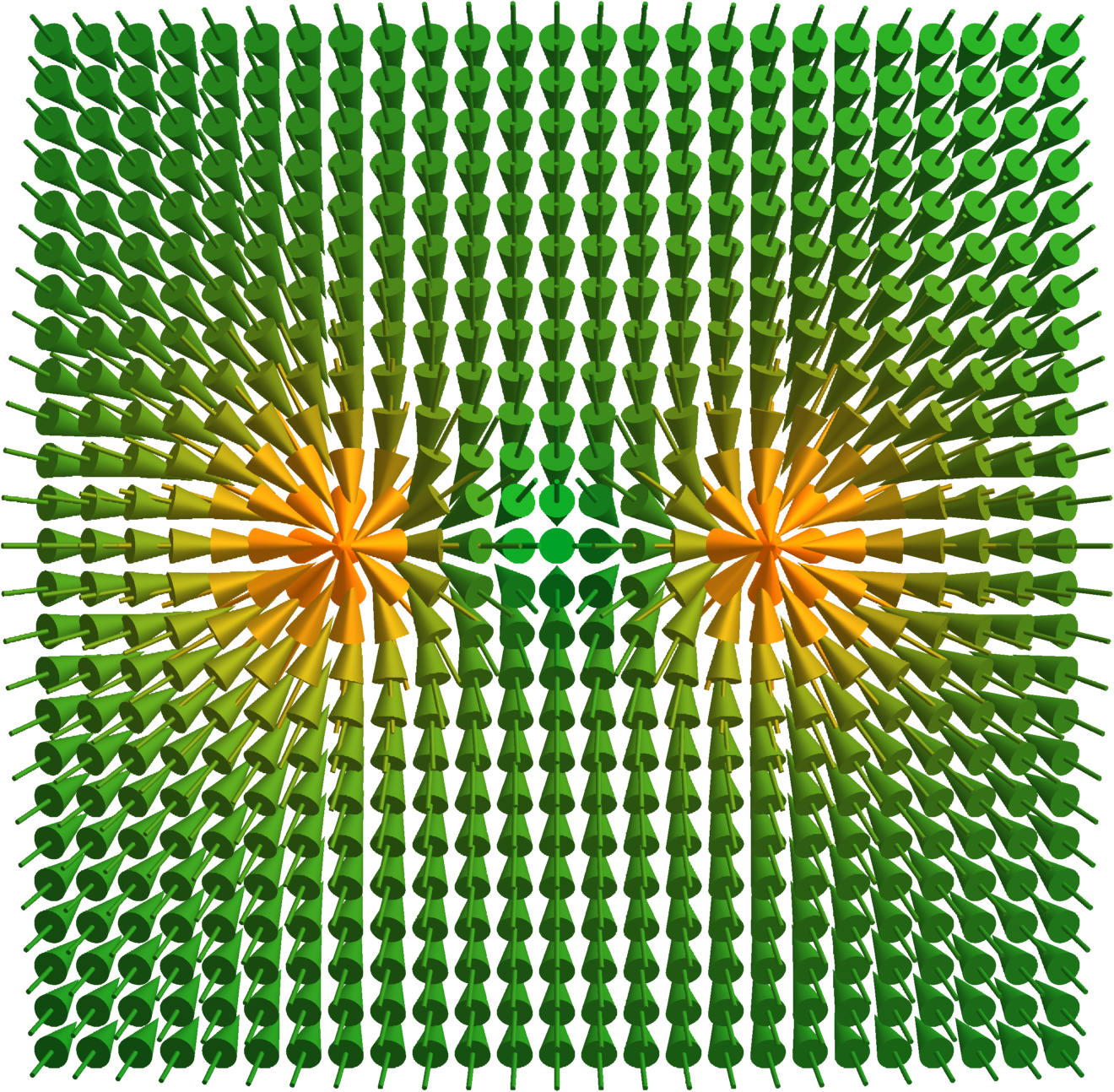}
\par\end{centering}
\caption{Two repelling outward Néel skyrmions.}

\label{Fig-Two_skyrmions}
\end{figure}

Skyrmions repel each other via two mechanisms. One is intrinsic repulsion
via the conflicting in-plane spin fields created by the two skyrmions.
This repulsion energy decreases exponentially on the distance $d$
between the skyrmions and has the form \citep{CGC-JPCM2020}
\begin{equation}
U(d)\simeq F\exp\left(-\frac{d}{\delta_{H}}\right),\qquad F\equiv60J\left(\frac{A^{2}}{JH}\right)^{2},\label{U_interaction}
\end{equation}
where $\delta_{H}=a\sqrt{J/|H|}$ is the magnetic length. The formula
above was obtained by fitting the numerical data for the skyrmion-skyrmion
repulsion energy and is valid in a wide range of $A$ and $H$. 

At large distances, the dominant interaction becomes the dipole-dipole
repulsion proportional to the number of magnetic layers in the film
and decaying as $1/d^{3}$, see Fig. 7 and Eq. (21) of Ref. \citep{CGC-JPCM2020}.
Here, we will take into account only the strong short-range interaction
and ignore the DDI repulsion.

\begin{figure}[ht]
\centering{}\includegraphics[width=8cm]{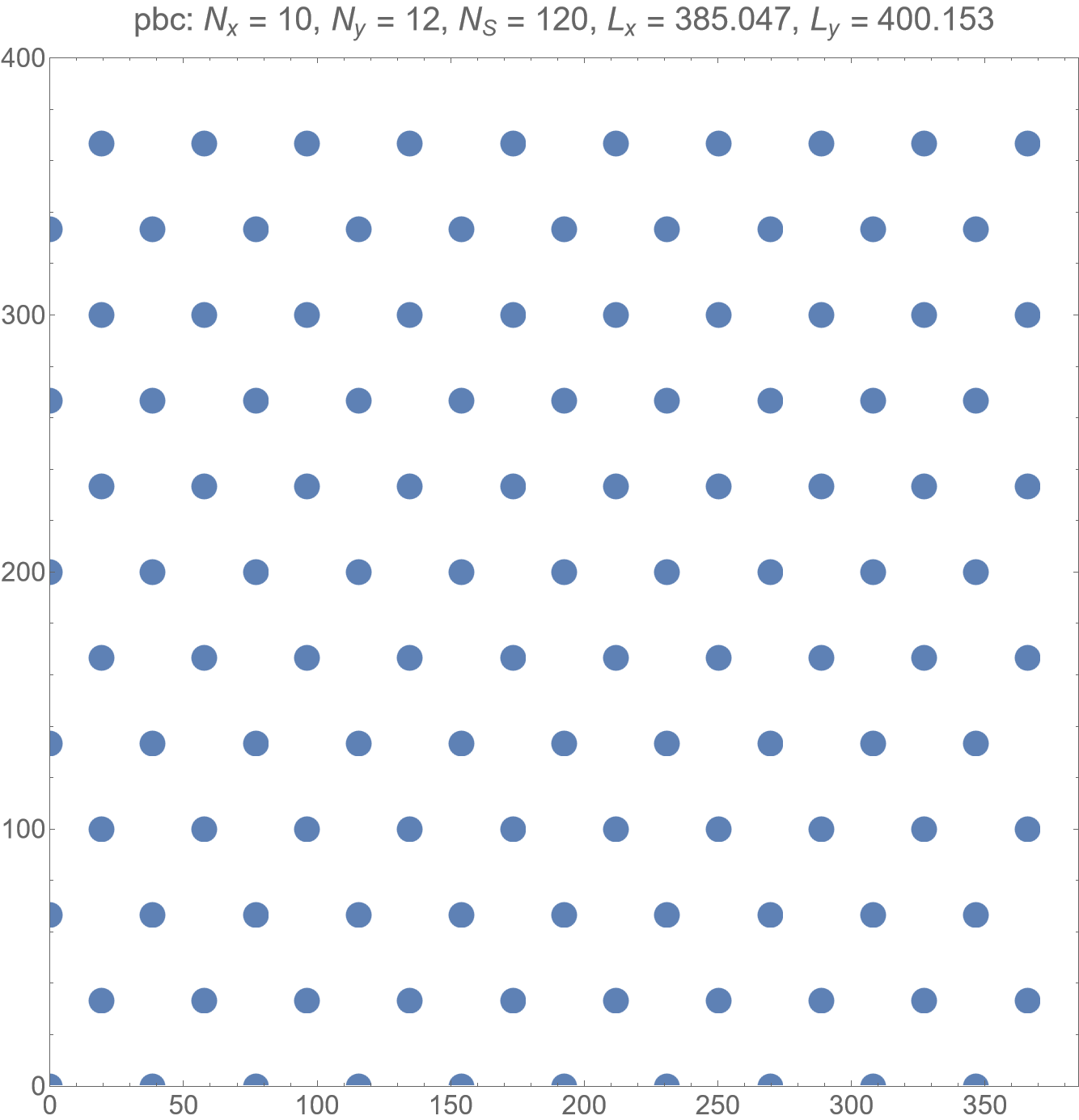}
\caption{Triangular lattice of skyrmions in a nearly square system with periodic
boundary conditions, $a_{S}=38.5a$.}
\label{Fig-TriangularLattice_pbc} 
\end{figure}

The number if skyrmions in the system is not fixed and can be found,
at least at low temperatures, from the minimization of the total system's
energy taking into account the skyrmion's core energy $\Delta E$
and the skyrmion-skyrmion interaction energy. Consider a system of
area $S$ containing a triangular skyrmion lattice of the period $a_{S}$,
see Fig.\ \ref{Fig-TriangularLattice_pbc}. The unit cell is a rhombus
of the area $a_{S}^{2}\cos60\lyxmathsym{\textdegree}=a_{S}^{2}\sqrt{3}/2$,
thus there are 
\begin{equation}
N_{S}=\frac{2}{\sqrt{3}}\frac{S}{a_{S}^{2}}
\end{equation}
skyrmions in the system. Each skyrmion in the lattice interacts with
its six nearest neighbors, whereas, as we shall see, the interaction
with further neighbors is negligibly small. Half of the interaction
energy can be ascribed to each skyrmion in a pair. Thus, the energy
per skyrmion is
\begin{equation}
E_{0}=\varDelta E+3F\exp\left(-\frac{a_{S}}{\delta_{H}}\right).\label{E0_def}
\end{equation}
The equilibrium state is defined by the minimization of the total
system's energy $E_{\mathrm{tot}}=N_{S}E_{0}$ with respect to $a_{S}$
that leads to the transcendental equation
\begin{equation}
-\frac{\varDelta E}{3F}=\left(1+\frac{x}{2}\right)e^{-x},\qquad x\equiv\frac{a_{S}}{\delta_{H}}.\label{aS_Eq}
\end{equation}
This transcendental equation has a solution for 
\begin{equation}
0<-\varDelta E/(3F)<1\label{aS_solution_condition}
\end{equation}
 that is possible only for the negative skyrmion's core energy $\varDelta E$.
For $\varDelta E\rightarrow0$, Eq. (\ref{aS_Eq}) yields $x\rightarrow\infty,$
then $E_{0}\rightarrow0$. 

Stability of the lattice requires $E_{0}<0$, that is, $-\varDelta E/(3F)e^{x}>1$.
Using Eq. (\ref{aS_Eq}), one can eliminate the exponential and rewrite
this condition as
\begin{equation}
-\frac{\varDelta E}{3F}e^{x}=1+\frac{x}{2}>1
\end{equation}
that is trivially satisfied with $x>0$. That is, if Eq. (\ref{aS_Eq})
has a solution, then the energy of the skyrmion lattice is always
negative, that is, it is below the energy of the uniform state. 

As an illustration, for our main set of parameters $A/J=0.2$ and
$H/J=-0.025$ one has $\varDelta E=-4.23J$, $\delta_{H}=6.32a$,
and $F=154J$, thus $-\varDelta E/(3F)=0.009156$ and the solution
of the transcendental equation yields $x=6.09$ and $a_{S}=38.5a$.
In this case, the nearest-neighbor interaction energy 
\begin{equation}
U_{0}=F\exp\left(-\frac{a_{S}}{\delta_{H}}\right)\label{U0_def}
\end{equation}
becomes $U_{0}=Fe^{-x}=0.3486J$. The next-nearest-neighbor interaction
energy is $U_{nnn}=Fe^{-\sqrt{3}x}=0.0040J$ and it is, indeed, negligible.

\begin{figure}
\begin{centering}
\includegraphics[width=9cm]{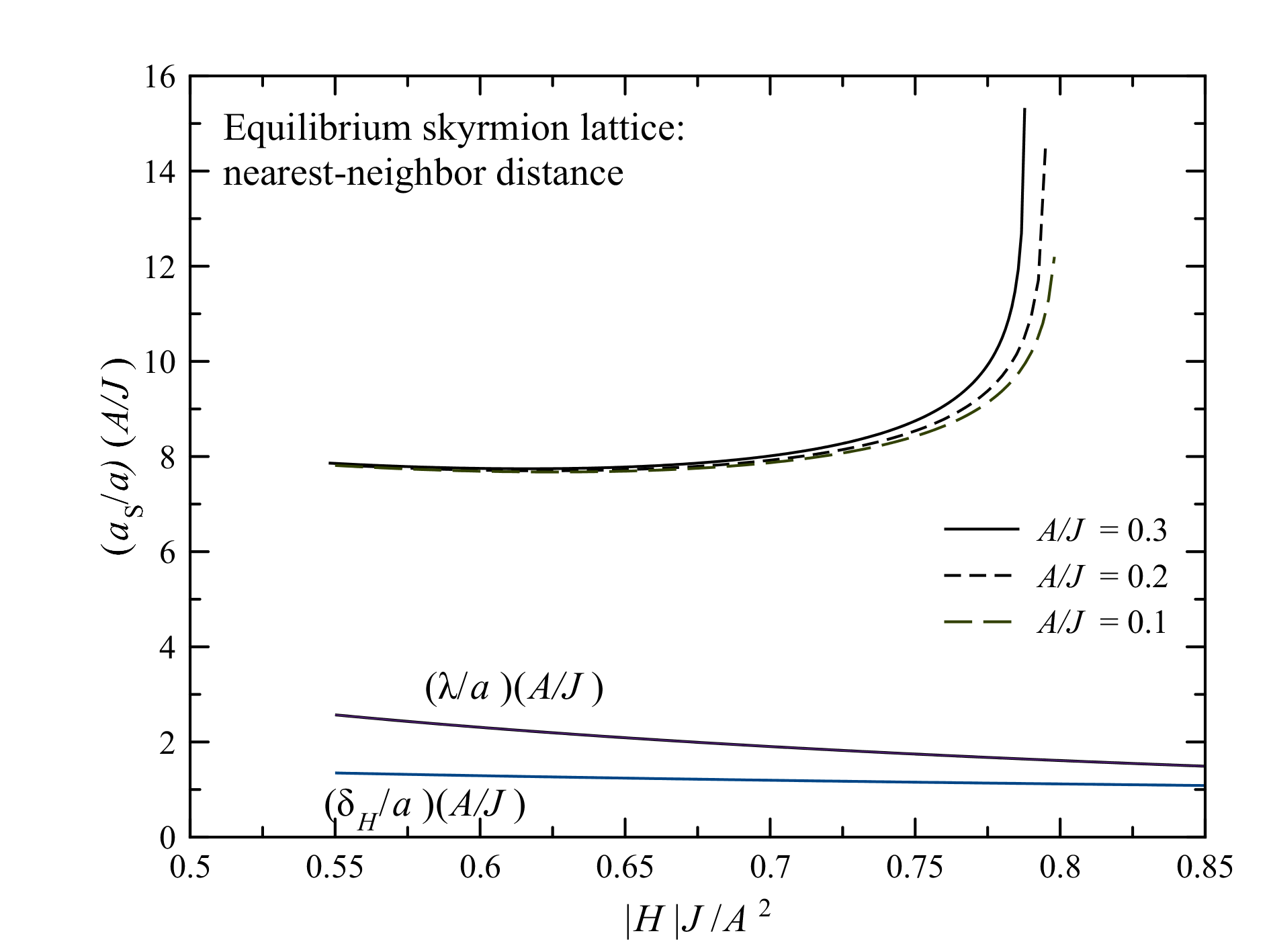}
\par\end{centering}
\caption{The equilibrium value of the skyrmion-lattice period $a_{S}$ vs $H$
for different values of $A$, compared with the skyrmion size $\lambda$
and the magnetic length $\delta_{H}$.}

\label{Fig-aS_vs_H_scaled_scaled}
\end{figure}

\begin{figure}
\begin{centering}
\includegraphics[width=9cm]{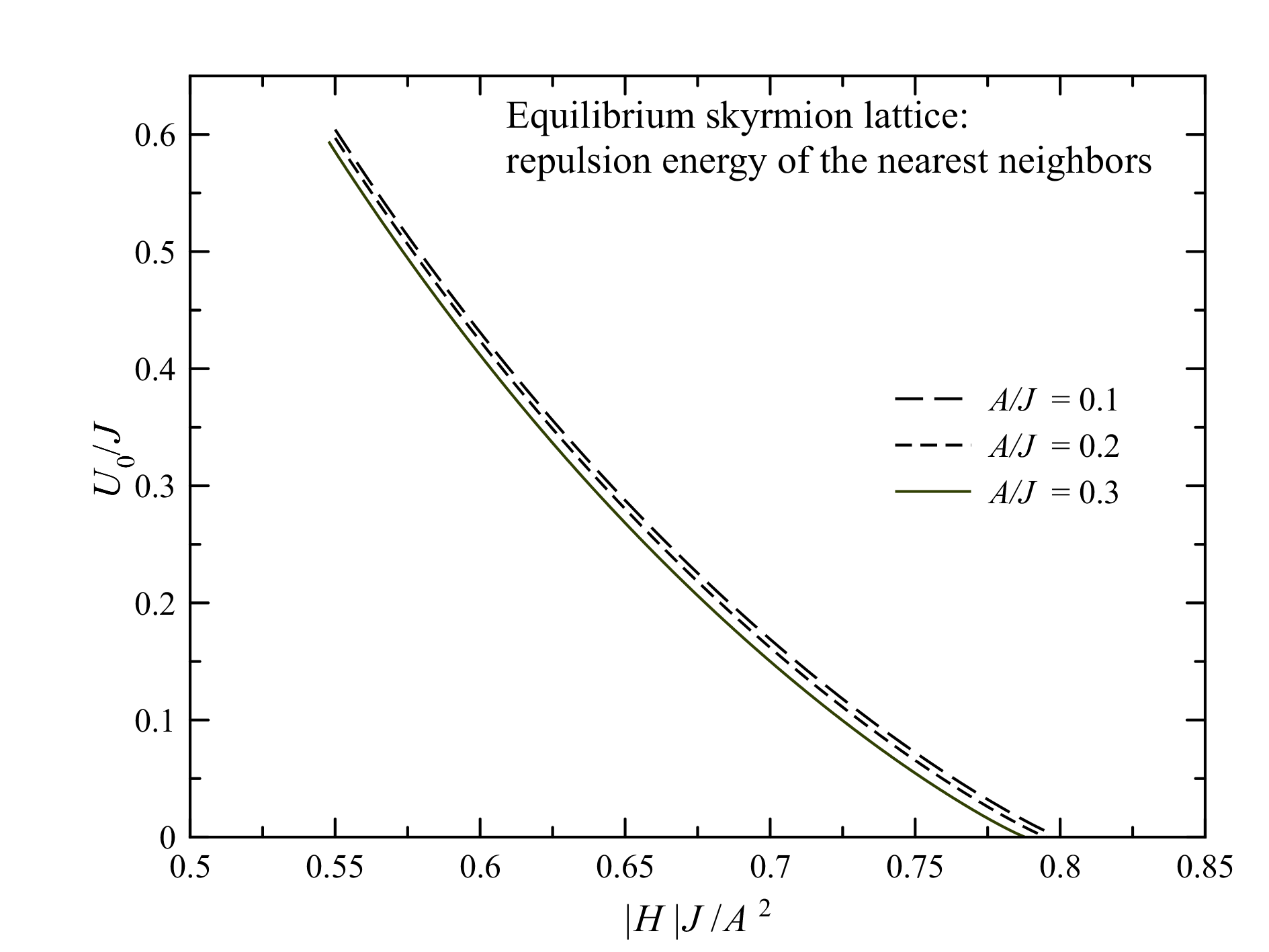}
\par\end{centering}
\caption{The nearest-neighbor repulsion energy $U_{0}$ in the equilibrium
skyrmion lattice vs $H$ for different $A$.}

\label{Fig-U0_vs_H_scaled}
\end{figure}
The energy of the equilibrium skyrmion lattice per skyrmion $E_{0}$
is shown by the blue lines in Fig. \ref{Fig-Delta_E_vs_H_scaled}
vs $H$ for different values of the DMI constant $A$. The values
of $E_{0}$ are negative, as it should be. The calculated values of
the skyrmion-lattice period $a_{S}$ are shown in Fig. \ref{Fig-aS_vs_H_scaled_scaled}.
In the whole parameter range where skyrmions are stable, $a_{S}\apprge\lambda$
that allows to consider skyrmions as point objects. Near $|H|J/A^{2}\simeq0.8$
where $\Delta E$ and $E_{0}$ vanish, $a_{S}$ is diverging. Finally,
the skyrmion-skyrmion interaction energy $U_{0}$ in the skyrmion
lattice is shown in Fig. \ref{Fig-U0_vs_H_scaled}. As near $|H|J/A^{2}\simeq0.8$
the distance between skyrmions becomes large, the skyrmion interaction
becomes small.

Although the number of skyrmions is not conserved and skyrmions can
be created and annihilated, these processes are very unlikely at the
temperatures near the melting point of the skyrmion lattice. For our
main parameter set $A/J=0.2$ and $H/J=-0.025$, removing one skyrmion
from the lattice increases the energy by $\Delta E_{-1}=-\Delta E-6U_{0}=2.14J$.
Adding an additional skyrmion at the optimal position in the middle
of any triangle increases the energy by $\Delta E_{1}=\Delta E+3F\exp(-r_{1}/\delta_{H})=9.48J$
if the interaction with the three nearest neighbors at the distance
$r_{1}=a_{S}/\left[2\cos(30\lyxmathsym{\textdegree})\right]=22.2a$
at the corner of the triangle are taken into account. If the three
next-nearest neighbors at the distance $44.5a$ are taken into account,
one obtains $\Delta E_{1}=9.89J$. On the other hand, the melting
temperature is $T_{m}\simeq0.12J$, so that in the problem of melting
a skyrmion lattice the high-energy processes changing the number of
skyrmions can be neglected and the number of skyrmions can be kept
constant. The equilibrium number of skyrmions found in this section
establishes at higher temperatures when the temperature is lowered.

\subsection{Constructing skyrmion lattices }

We study skyrmion lattices in several typical geometries such as nearly
square system with periodic boundary conditions (pbc), nearly square
system with rigid walls, rhomboid system with rigid walls, and circular
system with rigid walls. Positions of skyrmions in the basic triangular
lattice are given by 
\begin{equation}
\frac{{\bf R}}{a_{S}}={\bf e}_{x}n_{x}+\mathbf{e}_{60}n_{60}={\bf e}_{x}\left(n_{x}+\frac{1}{2}n_{60}\right)+{\bf e}_{y}\frac{\sqrt{3}}{2}n_{60},\label{Triangular_lattice_basic}
\end{equation}
where ${\bf e}_{x}$ and ${\bf e}_{y}$ are unit vectors along $x,y$
coordinate axes, $\mathbf{e}_{60}=(1/2){\bf e}_{x}+(\sqrt{3}/2){\bf e}_{y}$
is the lattice vector directed at $60\lyxmathsym{\textdegree}$ to
$x$-axis and $n_{x}$, $n_{60}$, are integers. The regions of $n_{x},n_{60}$
and the dimensions of the system are chosen to avoid distortions of
the lattice near the boundaries that is possible only for the pbc
and rhombic system.

An example of the near-square pbc system is shown in Fig. \ref{Fig-TriangularLattice_pbc}.
In this case, it is convenient to use $n_{x},n_{y}$ values in the
intervals $1\leq n_{x}\leq N_{x}$ and $1\leq n_{60}\leq N_{y}$ with
$N_{y}=2\mathrm{Int}(N_{x}/\sqrt{3})+2$, where $N_{y}$ is the number
of rows in the lattice, and define the lattice points as
\begin{equation}
\frac{{\bf R}}{a_{S}}={\bf e}_{x}\left[n_{x}-1+\mathrm{Fr}\left(\frac{n_{y}-1}{2}\right)\right]+{\bf e}_{y}\frac{\sqrt{3}}{2}\left(n_{y}-1\right),\label{Triangular_lattice_pbc}
\end{equation}
where $\mathrm{Int}(x)$ and $\mathrm{Fr}(x)$ are integer and fractional
parts of $x$. The number of skyrmions in this lattice is $N_{S}=N_{x}N_{y}$.
The system dimensions are chosen as
\begin{equation}
L_{x}=a_{S}N_{x},\qquad L_{y}=a_{S}\frac{\sqrt{3}}{2}N_{y}.\label{System_sizes_pbc}
\end{equation}
For large systems one has $L_{x}\cong L_{y}$ and the shape is close
to a square. One can see in Fig. \ref{Fig-TriangularLattice_pbc}
that the system has a smooth periodicity in $x$- and $y$-directions.

\begin{figure}
\begin{centering}
\includegraphics[width=6cm]{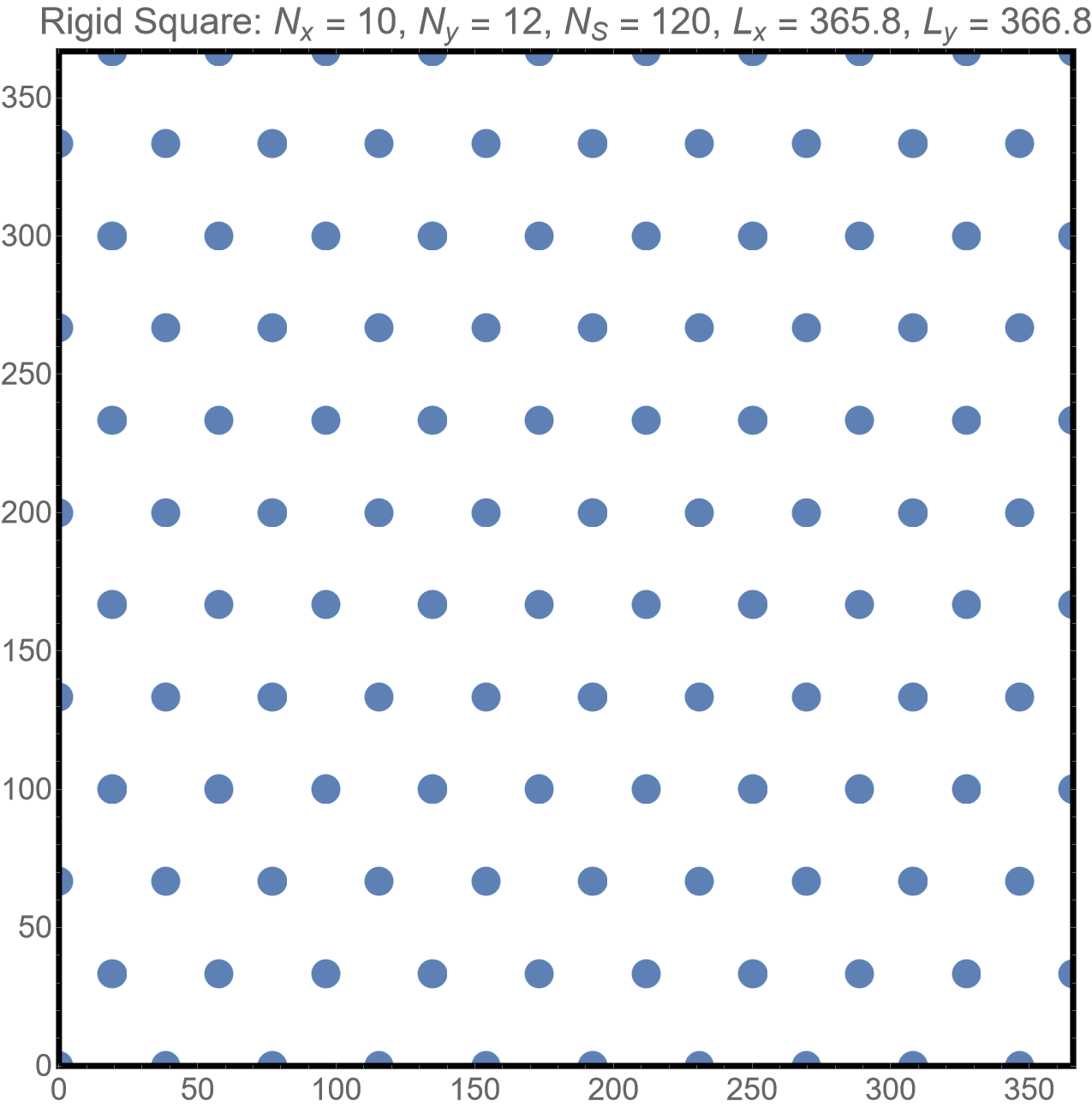}
\par\end{centering}
\begin{centering}
\includegraphics[width=8cm]{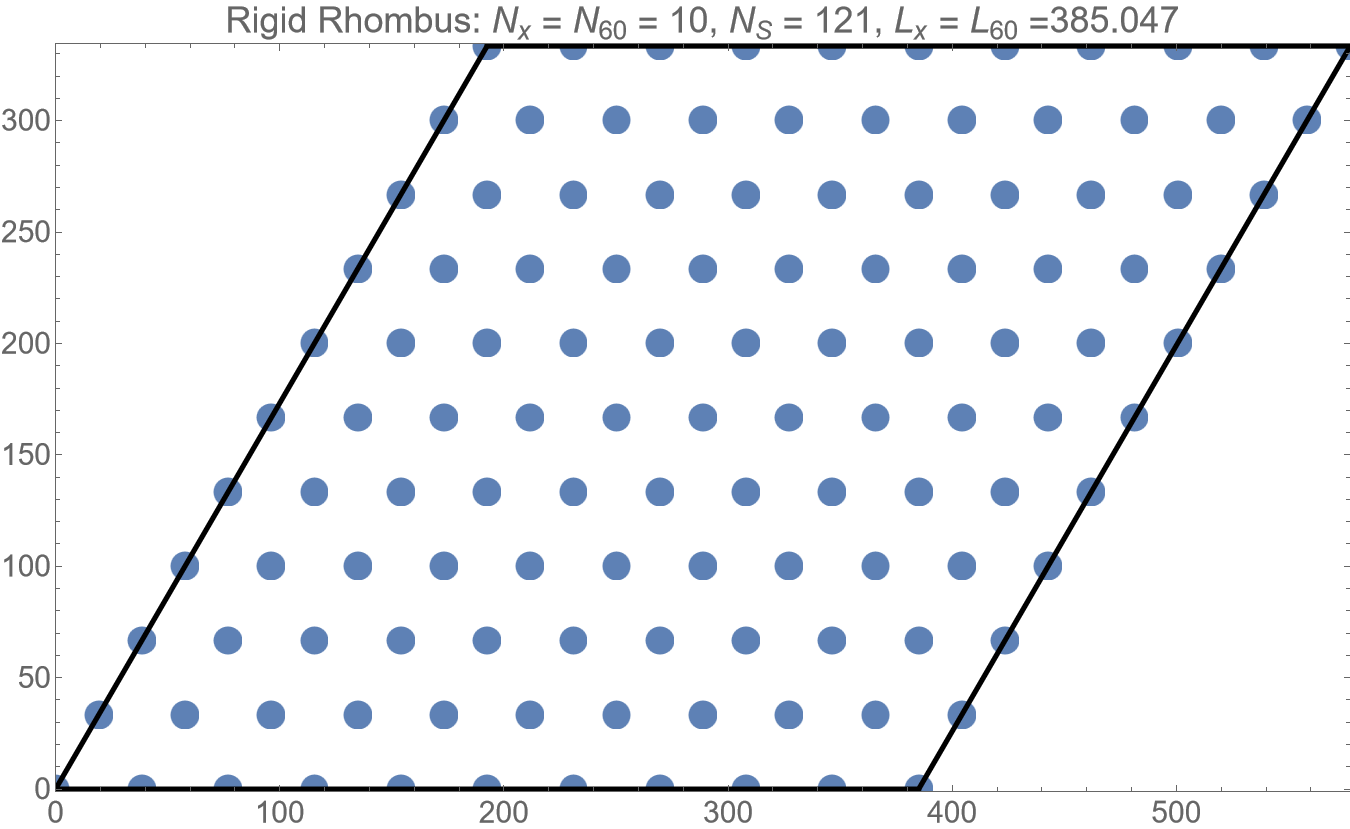}
\par\end{centering}
\caption{Triangular lattice of skyrmions in a nearly square and rhomboid systems
with rigid boundaries.}

\label{Fig-Triangular_Lattice_rigid}
\end{figure}

A near-square triangular lattice with rigid boundaries is constructed
in a similar way with minor modifications, see Fig. \ref{Fig-Triangular_Lattice_rigid}
(upper). Equation (\ref{Triangular_lattice_pbc}) and the definitions
of $N_{x}$ and $N_{y}$ remain the same, while the system dimensions
are given by 
\begin{equation}
L_{x}=a_{S}\left(N_{x}-\frac{1}{2}\right),\qquad L_{y}=a_{S}\frac{\sqrt{3}}{2}\left(N_{y}-1\right).\label{System_sizes_rigid_boundaries}
\end{equation}
In this system, the lattice becomes distorted near the vertical walls
as the skyrmion in all rows will be pressed to the walls by the internal
pressure resulting from the skyrmion repulsion.

For the system of a rhomboid shape with rigid boundaries, see Fig.
\ref{Fig-Triangular_Lattice_rigid} (lower), we use Eq. (\ref{Triangular_lattice_basic})
with $0\leq n_{x},n_{60}\leq N_{x}$, so that there are $N_{S}=\left(N_{x}+1\right)^{2}$
skyrmions. The system dimensions are $L_{x}=L_{60}=a_{S}N_{x}$. In
this case, the internal pressure is not distorting the lattice.

For the system of a circular shape, the boundaries favor different
orientations of the lattice cells at different positions near the
boundary that introduces frustration and makes a perfect triangular
lattice impossible. In this case the system crystallizes into a polycrystalline
state as the only possible scenario, see Fig. \ref{Fig-Circular_system}.
However, for other system shapes a polycrystalline state also emerges
on cooling, if the systems size is not too small.

\section{Physical quantities of triangular lattices}

\label{Sec-Physical-quantities}

In this section, we define physical quantities that characterize the
triangular lattice at zero and non-zero temperatures. Apart of the
energy, in the focus of interest are orientational and translational
orders.

\subsubsection{Orientational order parameter and correlation function}

The orientation of the hexagon of nearest neighbors of any particle
$i$ in the lattice is quantified by local hexagonality \citep{NH-PRB1979}
\begin{equation}
\Psi_{i}=\frac{1}{6}\sum_{j}\exp(6i\theta_{ij}),\label{Hexagonality_def}
\end{equation}
where the summation is carried out over 6 nearest neighbors $j$,
$\theta_{ij}$ is the angle between the $ij$ bond and any fixed direction
in the 2D lattice. If $\theta$ is counted from the direction of the
$x$-axis (which is our choice), and two of the sides of the perfect
hexagon coincide with the $x$-axis (the \textit{horizontal} hexagon
orientation), then all terms in the sum are equal one and $\Psi_{i}=1$.
For any other orientation of a perfect hexagon, $\Psi_{i}$ is a complex
number of modulus 1. In particular, the \textit{vertical }hexagon
orientation obtained by rotation by $30\lyxmathsym{\textdegree}$
from the horizontal orientation has $\Psi_{i}=-1$. The angle $\phi_{i}$
by which the hexagon $i$ is rotated from its initial horizontal orientation
is related to the phase angle $\Theta_{i}$ in $\Psi_{i}=\left|\Psi_{i}\right|e^{i\Theta_{i}}$
by $\phi_{i}=\Theta_{i}/6$. At finite temperatures, orientations
of the bonds fluctuate and the condition $|\Psi_{i}|=1$ no longer
holds. One can introduce the hex value averaged over the system,
\begin{equation}
V_{6}=\sqrt{\left\langle |\Psi_{i}|^{2}\right\rangle _{i}}\label{V6_def}
\end{equation}
that describes the quality of hexagons. At high temperature, when
the order is completely destroyed, the orientations of the bonds and
the angles $\theta_{ij}$ together with them become random. In this
limit $\left\langle |\Psi_{i}|^{2}\right\rangle _{i}=1/6$ as each
particle has six nearest neighbors on average. To describe the common
orientation of hexagons in the lattice, one can introduce the orientational
order parameter. 
\begin{equation}
O_{6}=\langle\Psi_{i}\rangle_{i}.\label{O6_def}
\end{equation}
Harmonic theory of 2D lattices yields a linear temperature dependence
of $O_{6}$ at low temperatures \citep{DC-PRB1995}. As we will see,
this linear decrease is due to the local disordering of individual
hexagons described by $V_{6}$. It is thus more convenient to plot
$O_{6}/V_{6}$ that describes the pure orientational order and only
weakly depends on temperature at low $T$.

In Eq. (\ref{Hexagonality_def}) one can define nearest neighbors
as those within a circle of radius $R_{0}$. A natural choice is $R_{0}=a_{S}\left(1+\sqrt{3}\right)/2$
that is the average between the nearest-neighbor distance and next-nearest-neighbor
distance. 

The orientational correlations in the lattice are defined as 
\begin{equation}
C_{6,ij}=\frac{\mathrm{Re}\left(\Psi_{i}\Psi_{j}^{*}\right)}{\left\langle |\Psi_{i}|^{2}\right\rangle _{i}}.\label{C6_def}
\end{equation}
To obtain the orientational correlation function (CF) that depends
on the distance $r_{ij}\equiv\left|\mathbf{r}_{i}-\mathbf{r}_{j}\right|$,
one has to compute all $C_{6,ij}$ values and corresponding distances
in the system. Then one can put the data on the plot. However, for
a large system there are too many plotting points. A better solution
is to make a histogram by binning the distances $r_{ij}$ into the
intervals of order $a_{S}$ and averaging $C_{6,ij}$ within each
bin. We use the bin size exactly equal to $a_{S}$, although other
choices are possible, too. One can see that the correlation function
defined above, $C_{6}(r)$, is equal to 1 at $r=0$ that corresponds
to $i=j$. 

In the liquid or a polycrystalline state, $C_{6}(r)\rightarrow0$
at $r\rightarrow\infty$. In this case one can use the finite-size
value $O_{6}\propto1/\sqrt{N_{S}}$ to estimate the corresponding
correlation radius $R_{6}$. One has
\begin{equation}
\left|O_{6}\right|^{2}=\frac{1}{N_{S}^{2}}\sum_{ij}\Psi_{i}\Psi_{j}^{*}\Rightarrow\frac{1}{N_{S}}\int_{0}^{\infty}\frac{2\pi rdr}{a_{S}^{2}}C_{6}(r).
\end{equation}
For the exponential CF, $C_{6}(r)=\exp\left(-r/R_{6}\right)$, one
obtains
\begin{equation}
\frac{R_{6}}{a_{S}}=\left|O_{6}\right|\sqrt{\frac{N_{S}}{2\pi}}.
\end{equation}

\subsubsection{Cristallinity}

\begin{figure}[ht]
\centering{}\includegraphics[width=8cm]{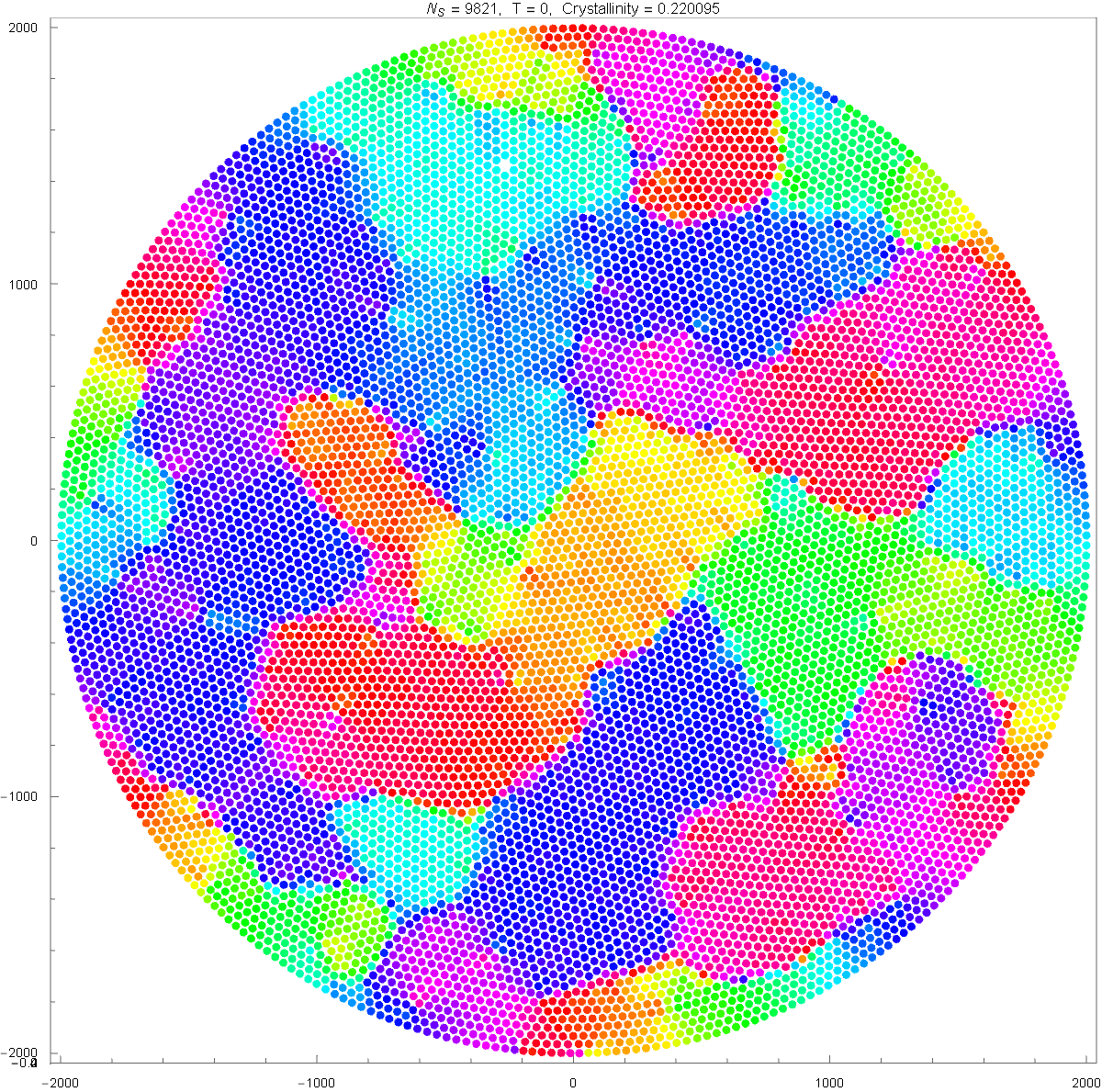} \caption{Polycrystalline structure of the skyrmion lattice in the magnetic
film of circular shape with $N_{S}=9821$ skyrmions at $T=0$, obtained
by cooling from high temperature.}
\label{Fig-Circular_system}
\end{figure}

Another measure of orientational order that we introduce here is \textit{cristallinity}.
Definition of crystallinity suggested below is based upon observation
that in the case of a total disorder the phase $\Theta_{i}$ angle
of $\Psi_{i}$ would be uniformly distributed in the interval $(-\pi,\pi)$.
Consequently, a sorted list of the values of $\Theta_{i}$ plotted
vs its index $i=1,2,...,N$ would be a straight line. Normalizing
$\Theta_{i}$ by $2\pi$ and $i$ by the number of particles $N$
produces a straight line going from $(0,-0.5)$ to $(1,0.5)$. When
differently oriented crystallites are present, the dependence of $\Theta_{i}$
on $i$ has plateaus in the ranges of $i$ that belong to the same
crystallite or similarly oriented crystallites, while the plateaus
are separated by narrow boundaries. This suggests definition of crystallinity
as the deviation from the straight line mentioned above: 
\begin{equation}
Crystallinity=\frac{4}{N}\sum_{i=1}^{N}\left|\Theta_{i}-\langle\Theta_{i}\rangle+\frac{i}{N}-0.5\right|,\label{Crystallinity_def}
\end{equation}
where $\langle\Theta_{i}\rangle$ is the average phase angle in the
system. This quantity can be easily computed after all $\Psi_{i}$
are found.

For a single crystal, $\Theta_{i}=\langle\Theta_{i}\rangle$, and
the sum in Eq.\ (\ref{Crystallinity_def}) reduces to summation of
the areas of two right-angle triangles, giving crystallinity of one.
If the system splits into $k$ crystallites of equal size, with equidistant
values of $\Theta$, crystallinity computed in a similar manner equals
$1/k$. This provides an estimate for the average number of particles
in a crystallite $N_{C}$ using $N_{C}/N=1/k$. In fact, one can define
$N_{C}=Crystallinity\times N$. 

It is convenient to show the polycrystalline state of a triangular
lattice with color coding using the function $\mathrm{Hue}\left(\Theta_{i}/(2\pi)\right)$.
As the argument of this function changes from 0 to 1, the color changes
as red, yellow, green, cyan, blue, magenta, and red again. An example
of a polycrystalline state in such a representation for a system of
circular shape obtained by cooling to $T=0$ from a completely disordered
state at $T=\infty$ is shown in Fig. \ref{Fig-Circular_system}.
Figs. \ref{Fig-TriangularLattice_pbc} and \ref{Fig-Triangular_Lattice_rigid}
could be color coded as red.

\subsubsection{Translational order}

The structure of the lattice is commonly described by the structure
factor 
\begin{equation}
S({\bf q})=\sum_{i}\exp(i{\bf q}\cdot{\bf r}_{i}).
\end{equation}
For a perfect lattice, $\left|S({\bf q})\right|$ has sharp maxima
at ${\bf q}$ equal to one of the reciprocal lattice vectors or their
linear combinations. For the triangular lattice with horizontal hexagons
($\Theta_{i}=0$) there are three reciprocal-lattice vectors
\begin{equation}
\mathbf{q}_{1}=(0,1)q,\quad\mathbf{q}_{2}=\frac{(\sqrt{3},-1)}{2}q,\quad\mathbf{q}_{3}=\frac{(-\sqrt{3},-1)}{2}q,\label{Reciprocal-lattice_vectors}
\end{equation}
where $q\equiv4\pi/\left(\sqrt{3}a_{S}\right)$. One can check ${\bf q}_{\nu}\cdot{\bf R}_{i}=2\pi$
for $\nu=1,2,3$ and ${\bf R}_{i}$ being any particle position in
a perfect lattice. This gives an idea to define the translational
order parameter for a given configuration as 
\begin{equation}
O_{\mathrm{tr}}\equiv\frac{1}{3N_{S}}\sum_{\nu=1}^{3}\left|\sum_{i}\exp(i{\bf q}_{\nu}\cdot{\bf r}_{i})\right|\label{Otr_def}
\end{equation}
that takes the value 1 for a perfect lattice and smaller values in
the presence of defects or thermal excitations. To improve the statistic,
we sum over all three reciprocal-lattice vectors.

In the same vein, translational correlations can be described by 
\begin{equation}
C_{\mathrm{tr},ij}=\frac{1}{3}\sum_{\nu=1}^{3}\cos\left[{\bf q}_{\nu}\cdot\left({\bf r}_{i}-{\bf r}_{j}\right)\right].\label{Ctr_def}
\end{equation}
Then, one can define the translational correlation function $C_{\mathrm{tr}}(r)$
in the same way as the orientational CF: computing the above expression
for every $i$ and $j$ and binning the results. 

From harmonic theory of 2D lattices follows that at any nonzero temperature
there is no true long-range translational order as long-wavelength
fluctuations make the correlation function a power law: 
\begin{equation}
C_{\mathrm{tr}}(r)\propto\left(\frac{a_{S}}{r}\right)^{\eta},\qquad\eta\propto T\label{Ctr_power}
\end{equation}
(see, e.g., Ref. \citep{DC-PRB1995}). Correspondingly, the translational
order parameter is not a true order parameter as it decreases to zero
in the thermodynamic limit. Expressing $O_{\mathrm{tr}}$via the CF
above, one obtains
\begin{equation}
O_{\mathrm{tr}}^{2}=\frac{1}{N_{S}}\sum_{i}C_{\mathrm{tr}}({\bf r}_{i})\sim\left(\frac{a}{L}\right)^{\eta},\label{Otr_vs_L}
\end{equation}
where $L$ is the linear system size. 

The definition of the translational order above relies on the single-crystal
structure of the lattice. Before computing in Eqs. (\ref{Otr_def})
and (\ref{Ctr_def}), one has to find the reciprocal-lattice vectors.
In the polycrystalline state, each crystallite has its own set of
${\bf q}_{\nu}$, so that translational order in the whole system
cannot be defined. 

Even in a single-crystal state at small but nonzero temperatures,
translational order in 2D is washed out by long-wavelength fluctuations,
so that $C_{\mathrm{tr}}(r)$ decreases as a power of $r$ and $O_{\mathrm{tr}}$
slowly decreases with the system size (see below). Still, one can
treat $O_{\mathrm{tr}}$ as a quasi-order parameter.

\subsubsection{Lattice defects}

\label{subsec:Lattice-defects}

Lattice imperfections can be generated by frustrating boundaries,
see Fig. \ref{Fig-Circular_system}, thermal excitations, etc. The
most common ones are boundaries between crystallites (grain boundaries),
vacancies, dislocations and disclinations.

Disclinations are centered at the lattice sites having five or seven
nearest neighbors and they result in the hexagon orientations being
different at any different point away from the disclination center.
Disclination with $z=5$ can be seen as the result of cutting a wedge
from a crystal with subsequent deforming the crystal to close the
gap. Disclination with $z=7$ can be seen as the result of inserting
a wedge of additional particles into the system. Thus, a single disclination
is a topological object and it should severely reduce both orientational
and translational order in the system.

Dislocations are incomplete rows of particles that end at the dislocation
center. A single dislocation can severely reduce the translational
order but only slightly disturbs the orientational order. According
to the common view, dislocations can be born in tightly-bound pairs
by thermal agitation and then the pairs unbind at some temperature.
Further, according to the KTHNY scenario, an isolated dislocation
consists of a pair of disclinations that unbind at another still higher
temperature.

\begin{figure}
\begin{centering}
\includegraphics[width=8cm]{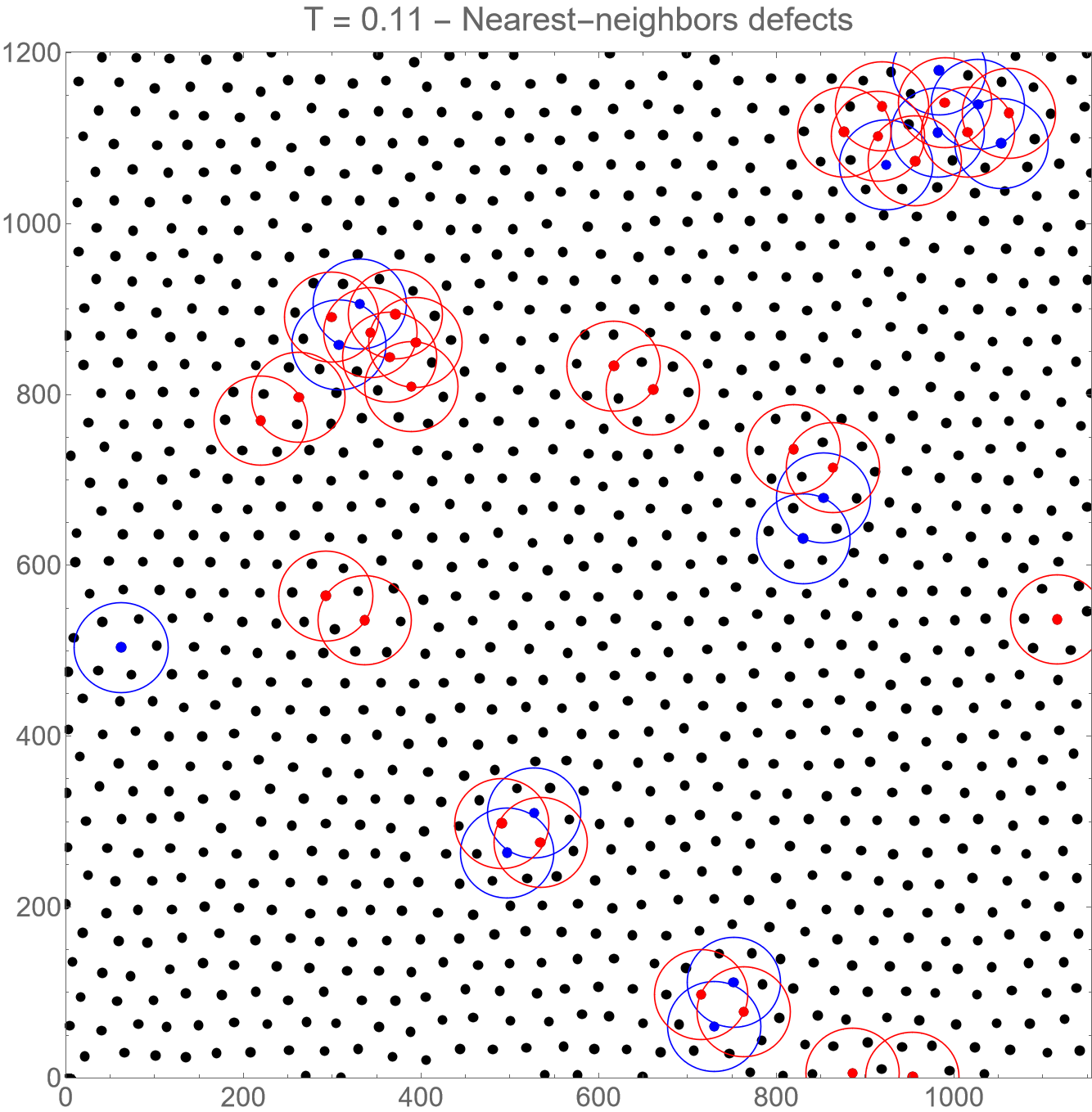}
\par\end{centering}
\begin{centering}
\includegraphics[width=8cm]{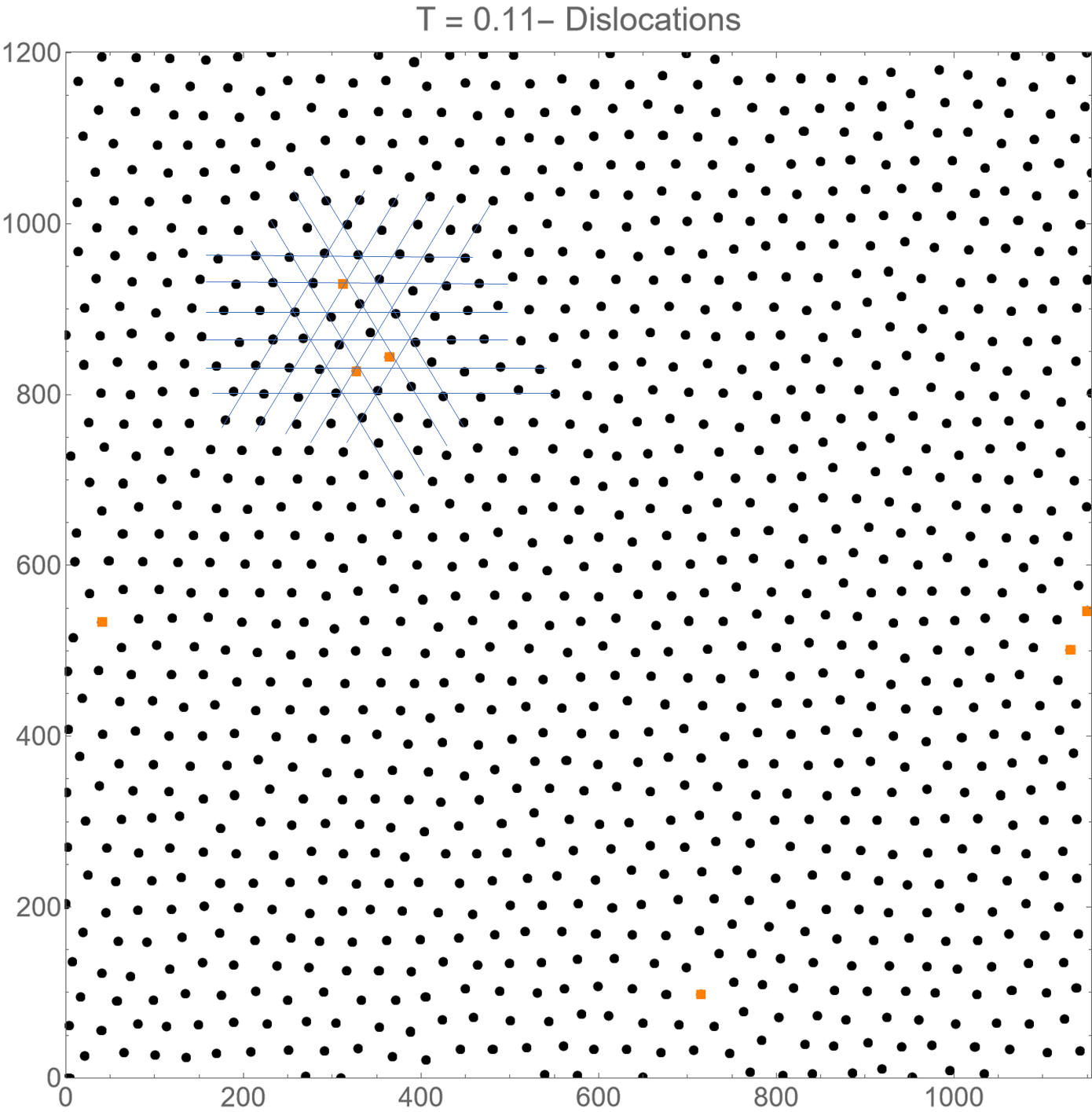}
\par\end{centering}
\caption{Skyrmion lattice at $T/J=0.11$. Upper panel: \textquotedblleft disclinations\textquotedblright{}
found by computing the number of nearest neighbors. Skyrmions with
5 and 7 nearest neighbors are marked by blue and red points and encircled
by blue and red circles of radius $R_{0}=a_{S}\left(\sqrt{3}+1\right)/2$.
Lower panel: \textquotedblleft dislocations\textquotedblright{} found
by computing the Burgers vector shown by orange squares.}

\label{Fig-SL_T}
\end{figure}

At $T=0,$ finding dislocations and disclinations is relatively straightforward.
Disclinations are found by counting the number of nearest neighbors
for each lattice site. Dislocations can be found by computing the
Burgers vector for each site. The numerical method is the following.
As a preliminary, for each lattice site $i$ the local orientation
of the lattice in its vicinity is found. For this,$\Psi_{j}$ is computed
for the site $i$ and all its neighbors within the cutoff radius.
For each of these sites, the phase angles $\Theta_{j}$ corresponding
to $\Psi_{j}$ are found and averaged. Dividing the result by 6, as
explained below Eq. (\ref{Hexagonality_def}), one obtains the average
orientation of the lattice in the vicinity of $i$. After that, a
minimal rhomboid trajectory around the site is constructed as follows.
First, the point expected as the bottom-left corner of the rhombus
in a perfect lattice is set and the particle closest to this point
is found by checking all candidates. Then the point shifted by one
lattice spacing in the positive $x$-direction from the found particle's
position is set and the particle closest to it is found. This step
is repeated to find the bottom-right corner of the rhomboid trajectory.
After that two similar steps are performed in the positive $60\lyxmathsym{\textdegree}$
direction to find the upper-right corner of the rhombus. Then two
steps are done in the negative $x$-direction to find the upper-left
corner of the rhombus. Finally, two steps are done in the negative
$60\lyxmathsym{\textdegree}$ direction to return to the bottom-left
corner of the rhombus. If the location of the latter coincides with
that found at the beginning, there is no dislocation. If one comes
to another site in the lattice, the Burgers vector is nonzero and
there is a dislocation. 

Whereas the procedure of finding dislocations and disclinations at
$T=0$ is straightforward, it become ambiguous at elevated temperatures
because of thermal shifting of particles from their equilibrium positions,
see Fig. \ref{Fig-SL_T} (upper). Local displacements of neighbors
of a site $i$ can result in the neighbor count different from six
without any change in the system's topology as a real disclination
does. Furthermore, the neighbor count at elevated temperatures substantially
depends on the choice of the range $R_{0}$ used for counting nearest
neighbors. A small increase of $R_{0}$ above the typical value $R_{0}=a_{S}\left(1+\sqrt{3}\right)/2$
substantially increases the number of sites with seven neighbors and
decreases the number of sites with five neighbors. This means that
the number of nearest neighbors is not well defined at elevated temperatures.
As the disclination is not a local but a global object, the method
of their identification has to be global, too. However, non of such
methods is currently known. 

Similarly, numerical identification of tightly bound dislocation pairs
at $T>0$ is unclear. Concerning isolated dislocations, one can compute
Burgers vectors as explained above but one can get false positives
because of too strong local thermal shifting of lattice sites from
their expected positions without any actual dislocation, see Fig.
\ref{Fig-SL_T} (lower). Ambiguities in identification of lattice
defects at elevated temperatures are also mentioned on page 4 of Ref.
\citep{Kapfer2015}. It seems that direct checking of the KTHNY scenario
of 2D melting by observing processes with dislocations and disclinations
is problematic.

\section{Numerical method}

\label{Sec-Numerical-method}

We compute the properties of the skyrmion lattice at nonzero temperatures
$T$, including melting, using the Metropolis Monte Carlo (MC) method.
All computations are performed for the set of parameters listed in
Sec. \ref{subsec:Skyrmion-skyrmion repulsion} {[}see above Eq. (\ref{U0_def}){]}.

The first thing to say is that the phase space of the system is very
complicated and there are valleys separated by high barriers, so that
thermally activated transitions between the valleys take a prohibitively
long time at low $T$, practically below the melting point. As an
example one can name the single-crystal state and polycrystalline
states, see Fig. \ref{Fig-Circular_system}. The Monte Carlo routine
reflects this property, that is, in general, it does not lead to averaging
over different valleys and the numerical results belong to a particular
valley. 

In the Metropolis Monte Carlo, particles $i$ are successively displaced
by a random vector $\Delta\mathbf{r}_{i}$ within the plane and the
change of the energy $\Delta E_{i}$ is computed. The move is accepted
if $\exp\left(-\Delta E_{i}/T\right)>\mathrm{rand}$, where rand is
a random real number in the range (0,1), otherwise it is rejected.
The trial displacement $\Delta\mathbf{r}_{i}$ should be essentially
smaller than the lattice period $a_{S}$ to prevent a one-move destruction
of the lattice. We used $\Delta\mathbf{r}_{i}$ in a random direction
with the length $\left|\Delta\mathbf{r}_{i}\right|=\mathrm{MCD}\,a_{S}\left(T/U_{0}+0.1\right)\mathrm{rand}$,
with the MC distance factor $\mathrm{MCD}=0.25$ and rand being a
random real number in the range (0,1). With $U_{0}=0.3486J$ and the
maximal temperature of the simulations $T_{\max}=0.2J$, the condition
$\left|\Delta\mathbf{r}_{i}\right|\ll a_{S}$ is fulfilled everywhere.
The average MC acceptance rate was about 0.65 at all temperatures
except the lowest ones.

MC simulations were performed with Wolfram Mathematica with compilation
in C on three different computers, the best of which, a Dell Precision
workstation, has 16 cores available for Mathematica. Simulations at
different $T$ were done in parallel cycles to maximize the usage
of computing resources and gain more statistics. For each temperature
in a cycle, we started with the same initial state using the perfect
lattice initial condition (LIC) or random initial condition (RIC).
The main body of computations was done for the system with pbc (see
Fig. \ref{Fig-TriangularLattice_pbc}) with $N_{x}=300$ and thus
the number of skyrmions $N_{S}=104400$. Another system size we used
was $N_{x}=100$ with $N_{S}=11600$. Some simulations were performed
for the nearly-square system with rigid walls {[}see Fig. \ref{Fig-Triangular_Lattice_rigid}
(upper){]} and same sizes. Also, we have done simulations for the
rhomboid system {[}see Fig. \ref{Fig-Triangular_Lattice_rigid} (lower){]}
and for the circular system (see Fig. \ref{Fig-Circular_system}).

The relaxation (thermalization) time near the melting point turns
out to be extremely long. For this reason, it is not feasible to perform
Monte Carlo simulations with a fixed number of Monte-Carlo steps (MCS).
Thus, one needs a MC routine with an automatic number of MCS defined
with the help of some stopping criterion. We looked at the evolution
of the system's energy $E$ that was computed with the interval of
$\mathrm{MCBlock}=10$ MCS. We define the running average of the energy
$E_{\mathrm{mean}}$ being the mean of the last $\mathrm{Round}(\mathrm{MCMass\,}N_{E})$
output values of energy, where $N_{E}$ is the total number of the
energy values and we used $\mathrm{MCMass}=0.3$ (that is, we averaged
the last 30\% of the obtained energy values). The value of $E_{\mathrm{mean}}$
that is much smoother than $E$ was used to formulate the stopping
criterion as follows. We define the MC span, typically $\mathrm{MCSpan}=300$,
and consider the last MCSpan values of the $E_{\mathrm{mean}}$ list.
Then we define the slope of the energy $E_{\mathrm{slope}}$ per MCS
using the averages of the first and second halves of the MCSpan list.
The stopping criterion used reads $\left|E_{\mathrm{slope}}\right|<\mathrm{MCSens}\,U_{0}$,
where MCSens is the MC sensitivity. The parameter $\mathrm{MCSens}$
takes very small values, from $10^{-7}$ down to $10^{-11}$. The
minimal number of MCS is $\mathrm{MCBlock}\times\mathrm{MCSpan}$,
typically 3000. The maximal number of MCS was set to $10^{6}$. To
prevent stopping in the case when $E_{\mathrm{slope}}$ simply changes
its sign, the stopping criterion was required to be fulfilled 10 times.

In many Monte Carlo simulations of the past, a large number of MCS
was done to reach equilibrium and then another large number of MCS
was used for the measurements of physical quantities in the equilibrium
state. The second stage is needed for small systems to average out
fluctuations. However, in large systems the measurement routine is
unnecessary since for most quantities (except, for instance, susceptibility
computed via the fluctuation formula) fluctuations within a particular
valley are suppressed by self-averaging while averaging over different
valley requires too much time and is not happening anyway. Thus for
a large system, one can use just the final state after stopping the
MC routine to compute physical quantities. Here, we used Monte Carlo
with and without measuring stage and we have seen the effect of the
measurement stage only for small systems above melting.

In the computation of the skyrmion-skyrmion interaction, we used the
cutoff distance $r_{\mathrm{cutoff}}=0.95\sqrt{3}a_{S}$ that is just
shy of the next-nearest distance $\sqrt{3}a_{S}$. For our set of
parameters, nearest-neighbors interaction is $U_{0}=0.3486J$ while
the interaction at the cutoff distance is $U_{\mathrm{cutoff}}=F\exp\left(-r_{\mathrm{cutoff}}/\delta{}_{H}\right)=0.00685J$
that is sufficiently small. Limiting the summation over interacting
partners can be done by two methods. 

First, for each particle, a list of particles within a range longer
than $r_{\mathrm{cutoff}}$ is maintained. The interaction energy
is computed by summation over these lists, whereas only the neighbors
within the range $r_{\mathrm{cutoff}}$ are taken into account. The
displacement of particles is monitored and the lists of prospective
partners are updated from time to time. This method works well below
melting as the particles remain near their initial places and the
lists of partners do not need to be updated. However, in the liquid
phase particles are traveling and the lists of partners have to be
updated more and more frequently as the temperature increases. In
fact, for large systems one needs to use several nested generations
of partners' lists as updating such list scanning the whole system
take much time. So, it is better to choose partners from a larger
partners' list. Maintaining high-order lists of partners of each particle
containing thousands of partners takes up a lot of memory. Thus, for
large systems above melting this method becomes problematic. An example
of a result obtained with the partner-lists method is Fig. \ref{Fig-Circular_system}.

Another method consists in splitting the system in regular bins, rectangular
for the the systems of rectangular shape and rhomboid for the rhomboid
system (see Fig. \ref{Fig-Triangular_Lattice_rigid}). Binning the
system, that is, finding the number and positions of particles in
each bin, is a fast procedure, and binning does not require too much
memory. To compute the interaction energy for each particle, one needs
to check for the prospective partners within the same bin and all
neighboring bins, in total, in nine bins. The smaller the bins, the
less partners are to check and the faster is the computation. However,
the sizes of the bins have to exceed $r_{\mathrm{cutoff}}$. This
method is slower below melting then the first one but it works better
overall. The binning procedure is performed after each MCS. The most
of the results are obtained by this second method.

Dislocations in the lattice are detected by computing the Burgers
vector for each skyrmion, as explained in Sec. \ref{subsec:Lattice-defects}.

\section{Numerical results}

\label{Sec-Numerical_results}

\begin{figure}
\begin{centering}
\includegraphics[width=9cm]{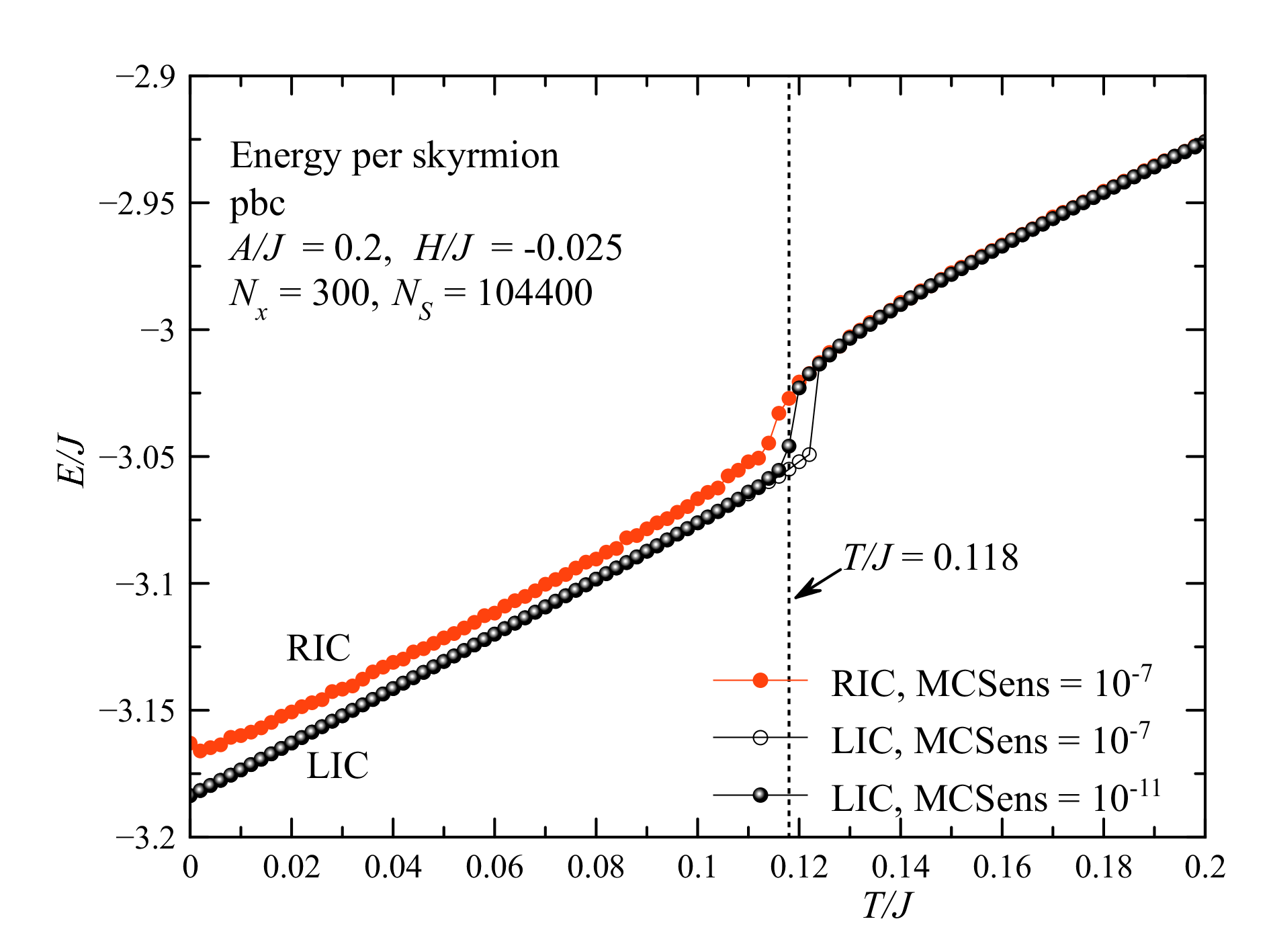}
\par\end{centering}
\caption{Temperature dependence of the energy per skyrmion $E$ for the lattice
and random initial conditions with different Monte Carlo sensitivities.}

\label{Fig-E_vs_T_pbc_Nx=00003D300}
\end{figure}

\begin{figure}
\begin{centering}
\includegraphics[width=9cm]{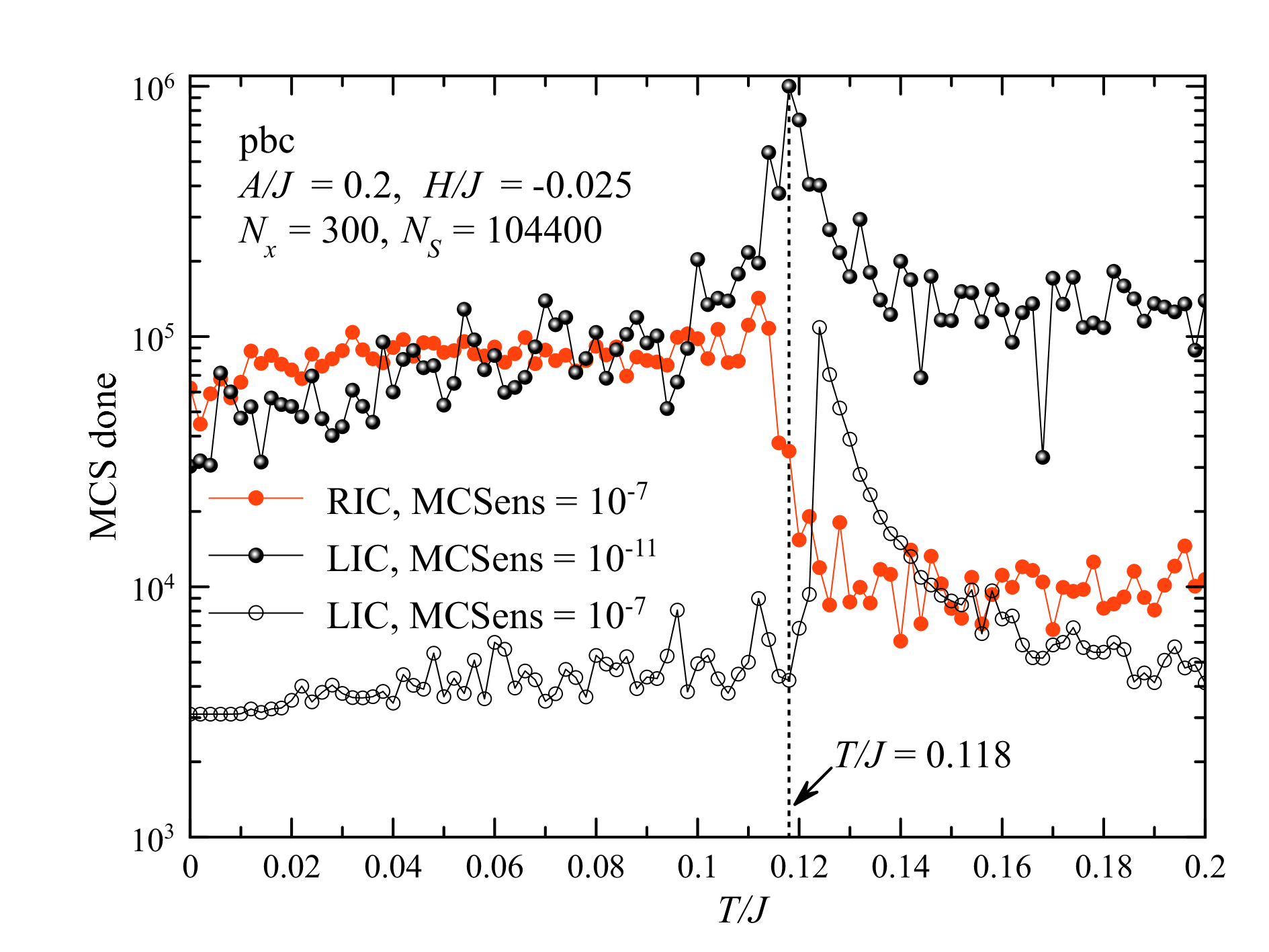}
\par\end{centering}
\caption{Temperature dependence of the number of Monte Carlo steps done. }

\label{Fig-MCS_vs_T_pbc_Nx=00003D300}
\end{figure}

\begin{figure}[ht]
\centering{}\includegraphics[width=9cm]{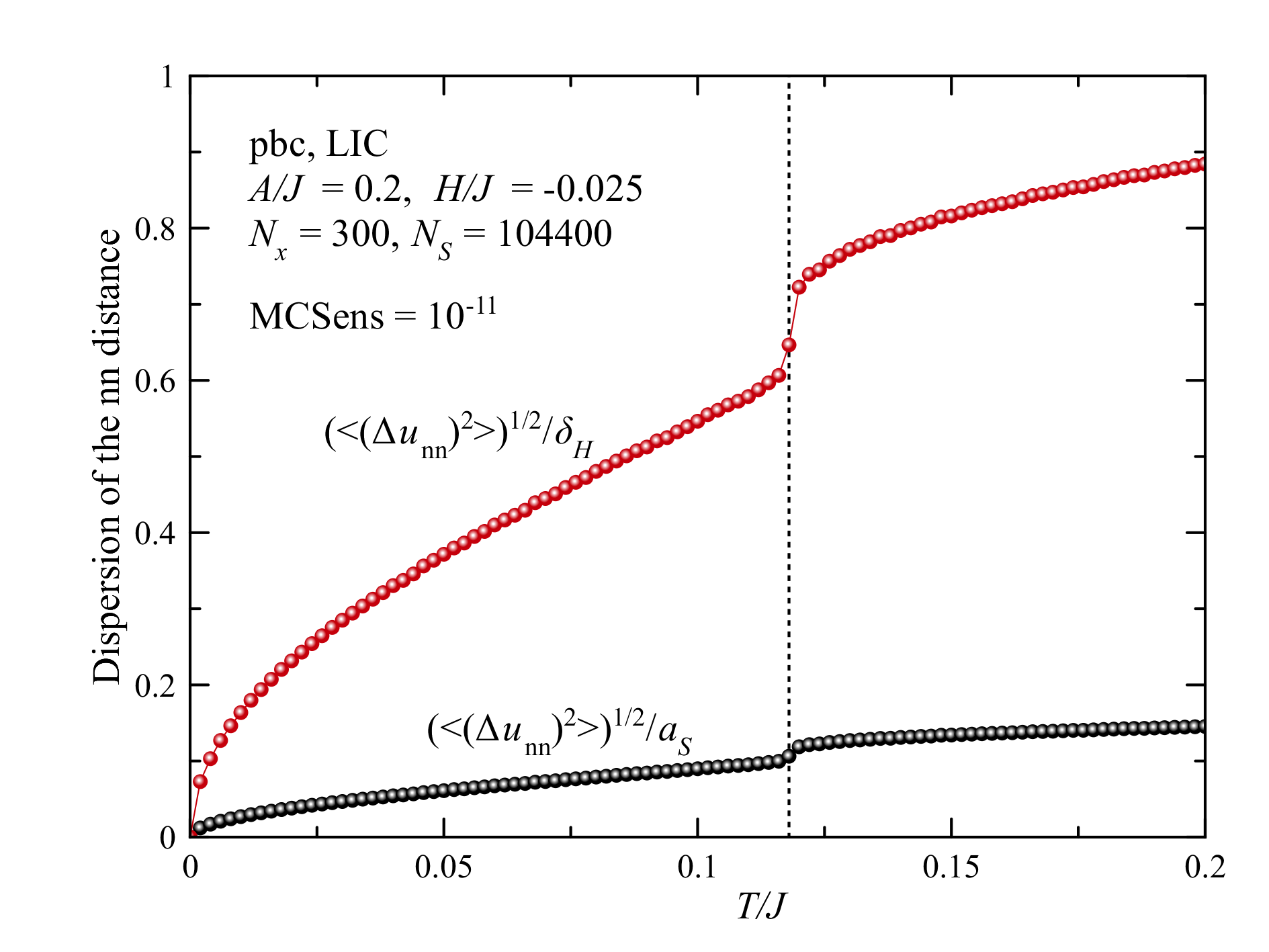}
\caption{Temperature dependence of the dispersion of the nearest-neighbor distance
in the skyrmion lattice.}
\label{Fig-Disp_of_nn_dist_vs_T_pbc_Nx=00003D300}
\end{figure}
\begin{figure}
\begin{centering}
\includegraphics[width=9cm]{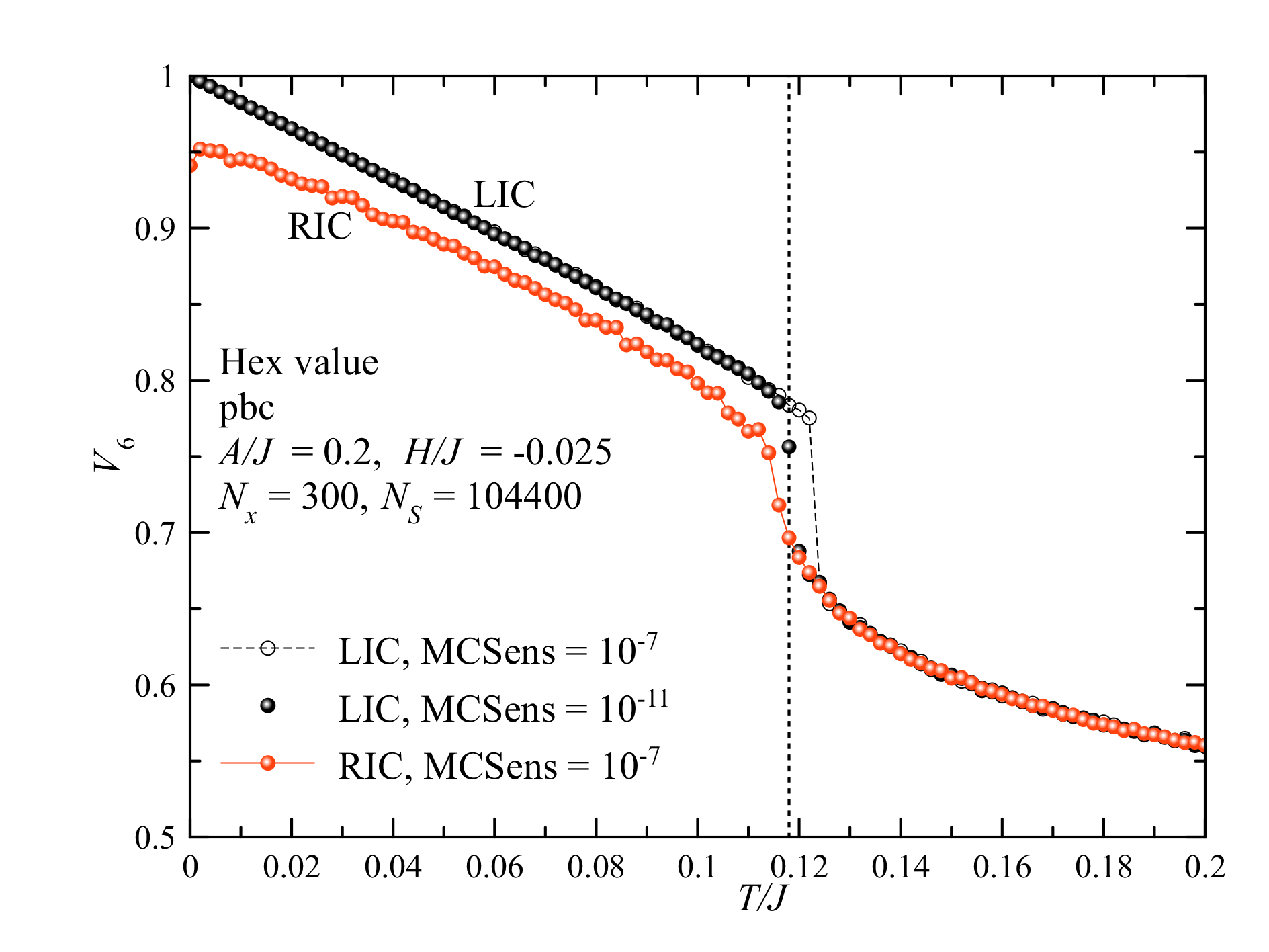}
\par\end{centering}
\caption{Temperature dependence of the average hexagonality value $V_{6}$,
Eq. (\ref{V6_def}). }

\label{Fig-Hex_value_vs_T_pbc_Nx=00003D300}
\end{figure}

\begin{figure}
\begin{centering}
\includegraphics[width=9cm]{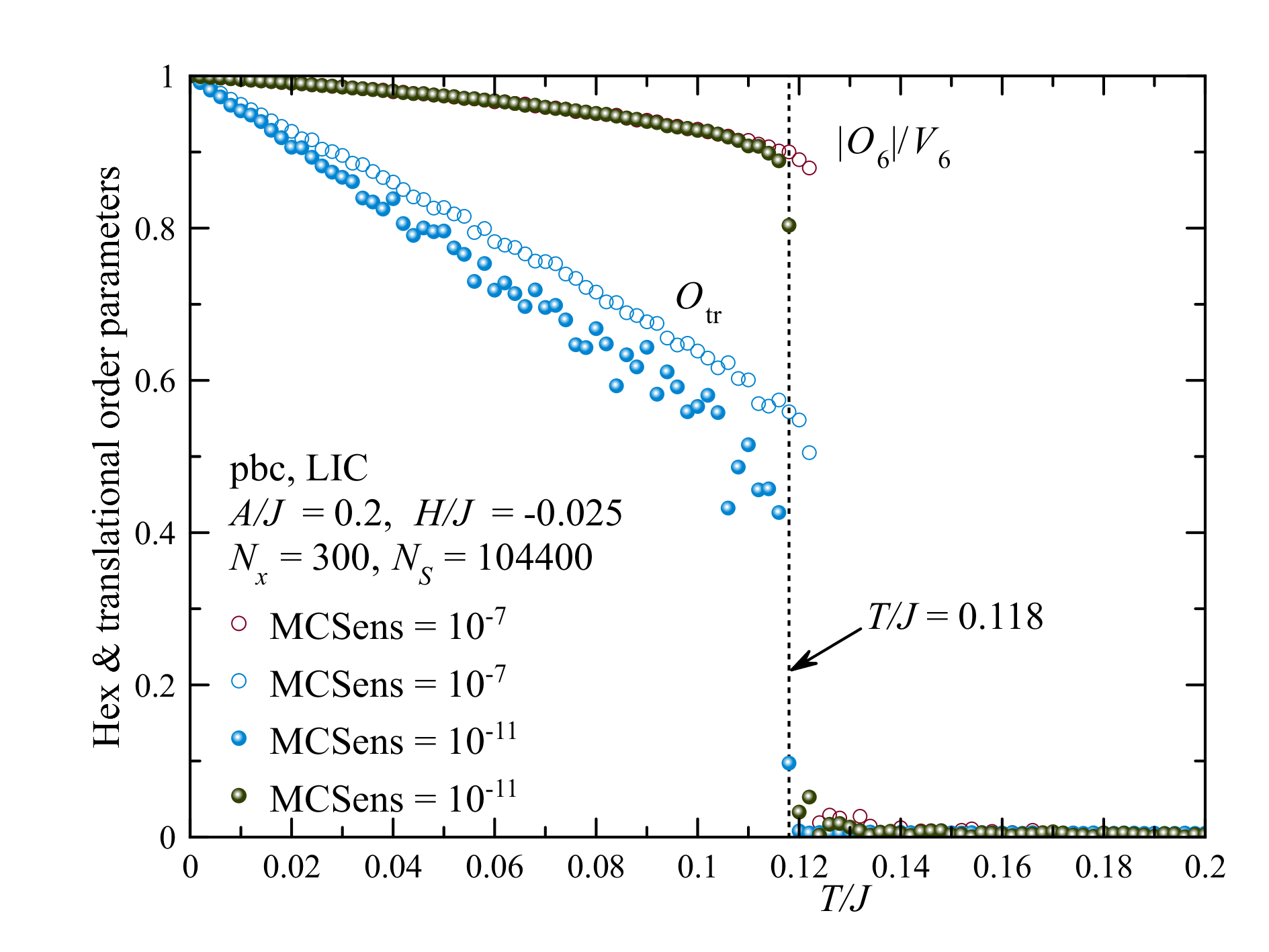}
\par\end{centering}
\caption{Temperature dependence of translational and orientational order parameters
in a nearly-square skyrmion lattice with pbc.}

\label{Fig-OrdPar_vs_T_pbc_Nx=00003D300}
\end{figure}
\begin{figure}
\begin{centering}
\includegraphics[width=9cm]{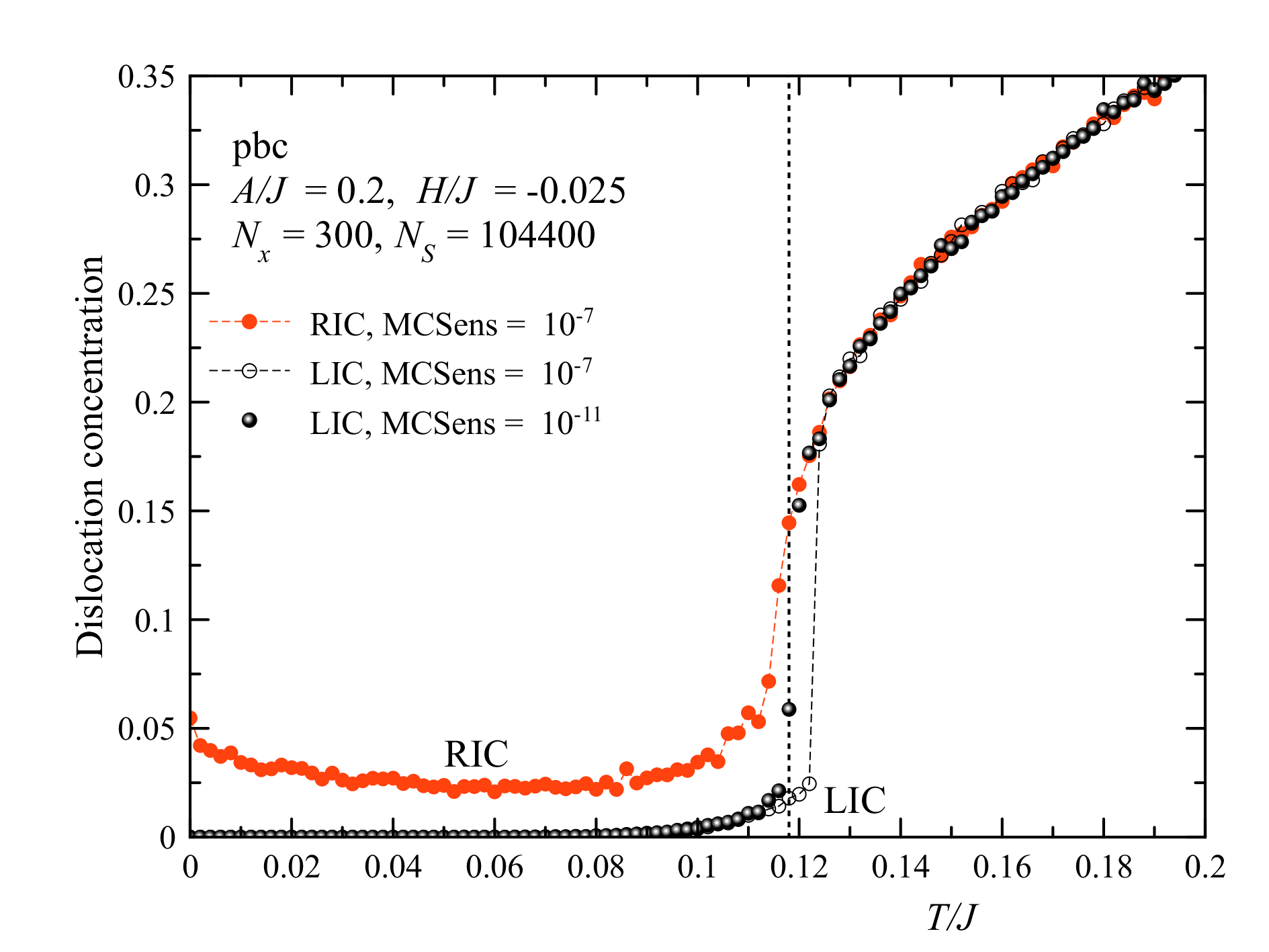}
\par\end{centering}
\caption{Temperature dependence of the concentration of dislocations, computed
via the Burgers vector.}

\label{Fig-NDisl_vs_T_pbc_Nx=00003D300}
\end{figure}

\begin{figure}
\begin{centering}
\includegraphics[width=9cm]{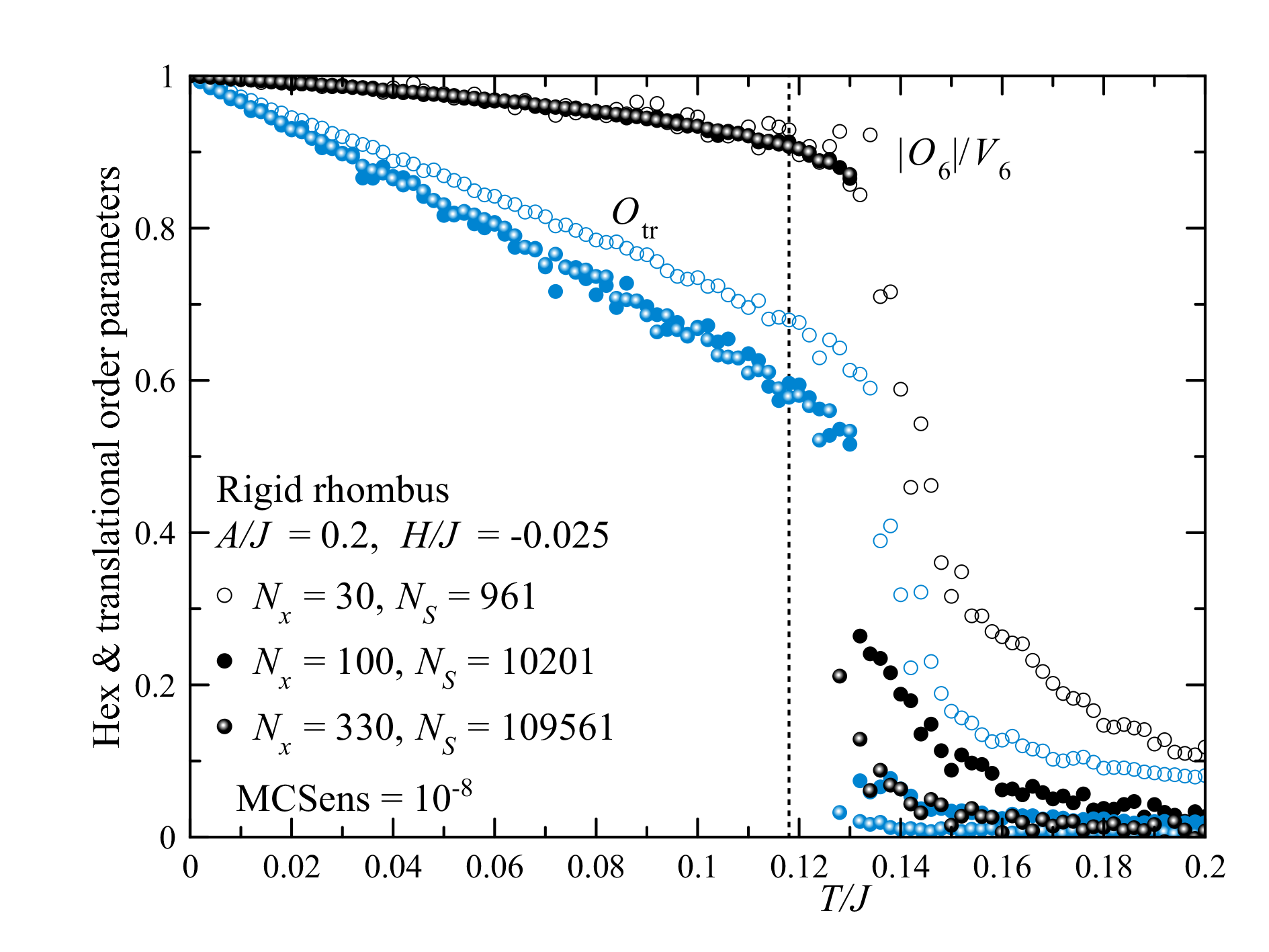}\label{Fig-OrdPar_vs_T_rhomb}
\par\end{centering}
\caption{Temperature dependence of translational and orientational order parameters
in a rhomboid skyrmion lattice with rigid boundaries, Fig. \ref{Fig-Triangular_Lattice_rigid}
(lower), with different system sizes. The dashed vertical line marks
the approximate melting temperature for the pbc system, $T/J=0.118$.}
\end{figure}

The main question in the melting of the skyrmion lattice is the same
as for all 2D systems: is the transition similar to that in 3D or
there are two successive transitions mediated by dislocation and disclinations
according to the KTHNY scenario? The results obtained show only one
transition that seems to be a first-order transition and has a hysteresis
with a big jump of physical quantities at the melting point. The position
of the jump on $T$ slightly depends on the length of the Monte Carlo
process, shifting to the left with the number of MCS done.

Temperature dependence of the order parameters for the system of rhomboid
shape with rigid boundaries, Fig. \ref{Fig-Triangular_Lattice_rigid}
(lower), is shown in Fig, \ref{Fig-OrdPar_vs_T_rhomb} for different
system sizes. This system shape favors the horizontal hexagon orientation
by the directions of its boundaries. The existence of this preferred
hexagon orientation is seen in the tail of $O_{6}$ in the liquid
phase. This tail does not exist in the model with pbc, Fig. \ref{Fig-OrdPar_vs_T_pbc_Nx=00003D300},
that does not set any preferred hexagon orientation. The tail is very
pronounced for small system sizes and weakens for larger systems.
This situation is resembling of a ferromagnetic system with a magnetic
field applied at the boundary. As one can see in the picture, rigid
boundaries also affect the translational order in finite-size systems,
inducing a partial ordering above the melting point. The tail of $O_{\mathrm{tr}}$
also weakens with the system size, as expected. The melting temperature
of the finite-size rhomboid system shifts to the right in comparison
with the pbc system because of the stabilizing effect of the boundaries.

\subsection{Temperature dependence of physical quantities in melting/freezing
of the skyrmion lattice}

The results for the three main systems studied here, near-square pbc
system, near-square rigid-boundary system, and rhomboid rigid-boundary
system are similar except for some subtleties that will be discussed
later. So, we show mainly the results of the pbc system with $\sim10^{5}$
skyrmions. The energy per skyrmion $E$ that contains the core energy
$\Delta E<0$ and the positive repulsion energy is shown in Fig. \ref{Fig-E_vs_T_pbc_Nx=00003D300}.
The energy values obtained starting from the perfect lattice (LIC)
for both sets of the MC parameters, \textbf{$\mathrm{MCSens}=10^{-7}$
}with $\mathrm{MCSpan}=300$ and $\mathrm{MCSens}=10^{-11}$ with
$\mathrm{MCSpan}=3000$ are the same everywhere except for the vicinity
of the melting point. Data for \textbf{$\mathrm{MCSens}=10^{-7}$
}show a jump at $T/J=0.122$ while the data \textbf{$\mathrm{MCSens}=10^{-11}$
}show a jump at $T/J=0.118$. This implies that melting is driven
by thermal activation over an energy barrier leading to nucleation
of the liquid phase, as in 3D. The energy values obtained starting
from random initial conditions (RIC) show a smaller jump at a lower
temperature, and the energy values below melting are higher than in
the case of LIC. The reason is that the system is freezing into a
polycrystal state, so that grain boundaries and other defects increase
the system's energy. Longer simulation can slightly reduce the energy
but formation of a single crystal in a large system required an exceedingly
large time. For realistic systems, formation of a single crystal will
be prevented by frustrating boundaries favoring different orientations
of the hexagons (see, e.g., Fig. \ref{Fig-Circular_system}) and by
pinning. 

Fig. \ref{Fig-MCS_vs_T_pbc_Nx=00003D300} shows the number of MCS
done before the thermalization routine stopped. For the LIC data,
thermalization at low temperatures is the shortest, with the numbers
of MCS about their minima $\mathrm{MCSpan}\times\mathrm{MCBlock}$,
that is, about $3000$ and 30000, respectively. Near the melting point,
thermalization becomes very slow, especially above melting where the
system's state changes significantly. For the second simulation, the
number of MCS reaches its cap of $10^{6}$ at $T/J=0.118$ without
fulfilling the stopping criterion. In the case of RIC, the relaxation
is very slow everywhere below freezing because of the slow motion
of grain boundaries.

Temperature dependence of the dispersion of the nearest-neighbor distance
in the skyrmion lattice $\sqrt{\left\langle \mathbf{u}_{nn}^{2}\right\rangle }$
is shown in Fig. \ref{Fig-Disp_of_nn_dist_vs_T_pbc_Nx=00003D300}.
It increases as $\sqrt{T}$ at low temperatures, as expected from
the theory of a harmonic lattice. There is a jump of the dispersion
at the melting point. The two curves are normalized by $a_{S}$ and
by the magnetic length $\delta_{H}$. The well-known empirical Lindemann
criterion predicts melting when the dispersion of the distances becomes
comparable with the lattice period. Practically, most of materials
melt at a lower temperature when the dispersion is about 0.1 of the
lattice period. Here, $\sqrt{\left\langle \mathbf{u}_{nn}^{2}\right\rangle }\simeq0.1a_{S}$
at the melting point. The dispersion normalized by $\delta_{H}$ is
of order one near melting. One can show this is the condition for
anharmonic terms in the expansion of the skyrmion lattice energy to
be comparable with the harmonic terms.

Fig. \ref{Fig-Hex_value_vs_T_pbc_Nx=00003D300} shows the temperature
dependence of the hex value $V_{6}$ defined by Eq. (\ref{V6_def}).
In the LIC case, it linearly decreases with $T$ at low temperatures
and jumps down at the melting point. Note that there is still a considerable
hexagonality in the liquid phase above melting. In the RIC case, $V_{6}$
is below one at low $T$ because of lattice defects.

Probably, the most interesting are the results for the rotational
and translational order parameters in the LIC case shown in Fig. \ref{Fig-OrdPar_vs_T_pbc_Nx=00003D300}.
The main (linear) temperature dependence of $O_{6}$, Eq. (\ref{O6_def}),
is due to the decrease of the hex value $V_{6}$, so we plot $\left|O_{6}/\right|V_{6}$
that describes the loss of hexagon's orientations with temperature.
This quantity goes quadratically at low $T$ and remains large until
melting, where it jumps down to zero. The curves for $\mathrm{MCSens}=10^{-7}$
and $10^{-11}$ are in a good agreement everywhere except the vicinity
of the melting point. 

To the contrast, the corresponding results for the translational order
parameter $O_{\mathrm{tr}}$, Eq. (\ref{Otr_def}), differ significantly.
While the $\left|O_{6}/\right|V_{6}$ are smooth below melting, those
for $O_{\mathrm{tr}}$ show a large scatter. This confirms that $O_{\mathrm{tr}}$,
unlike $O_{6}$, is not a good thermodynamic order parameter. There
is only a translational quasi-order smeared by long-wavelength fluctuations
and going to zero in the thermodynamic limit. Here we show that for
a large but finite system $O_{\mathrm{tr}}$ exists but is difficult
to compute as its convergence with the number of MCS is very slow.
Also, $O_{\mathrm{tr}}$ is extremely sensitive to lattice defects,
and a single dislocation or disclination in a large system can strongly
reduce its value. Note that a longer thermalization creates more scatter
in $O_{\mathrm{tr}}$, apparently due to thermal creation of lattice
defects that is a slow process.

Finally, the temperature dependence of the concentration of dislocations
$N_{\mathrm{disl}}/N_{S}$ computed in the same simulations via the
Burgers vector as explained in Sec. \ref{subsec:Lattice-defects}
is shown in Fig. \ref{Fig-NDisl_vs_T_pbc_Nx=00003D300}. In the LIC
case, the concentration of dislocations is very small at low temperatures,
then increases and makes an upward jump at the melting point. Above
melting it continues to increase and takes values of order one that
should be characteristic of a liquid state. In the RIC case, the concentration
of dislocations does not go to zero at $T\rightarrow0$. These results
should be taken with a grain of salt. As explained in Sec. \ref{subsec:Lattice-defects},
there should be false positives because of local lattice distortions
that are not real dislocations. Similarly, it is easy to compute the
concentration of skyrmions having 5 or 7 neighbors but at elevated
temperatures it does not prove the existence of disclinations.

\begin{figure}[ht]
\centering{}\includegraphics[width=9cm]{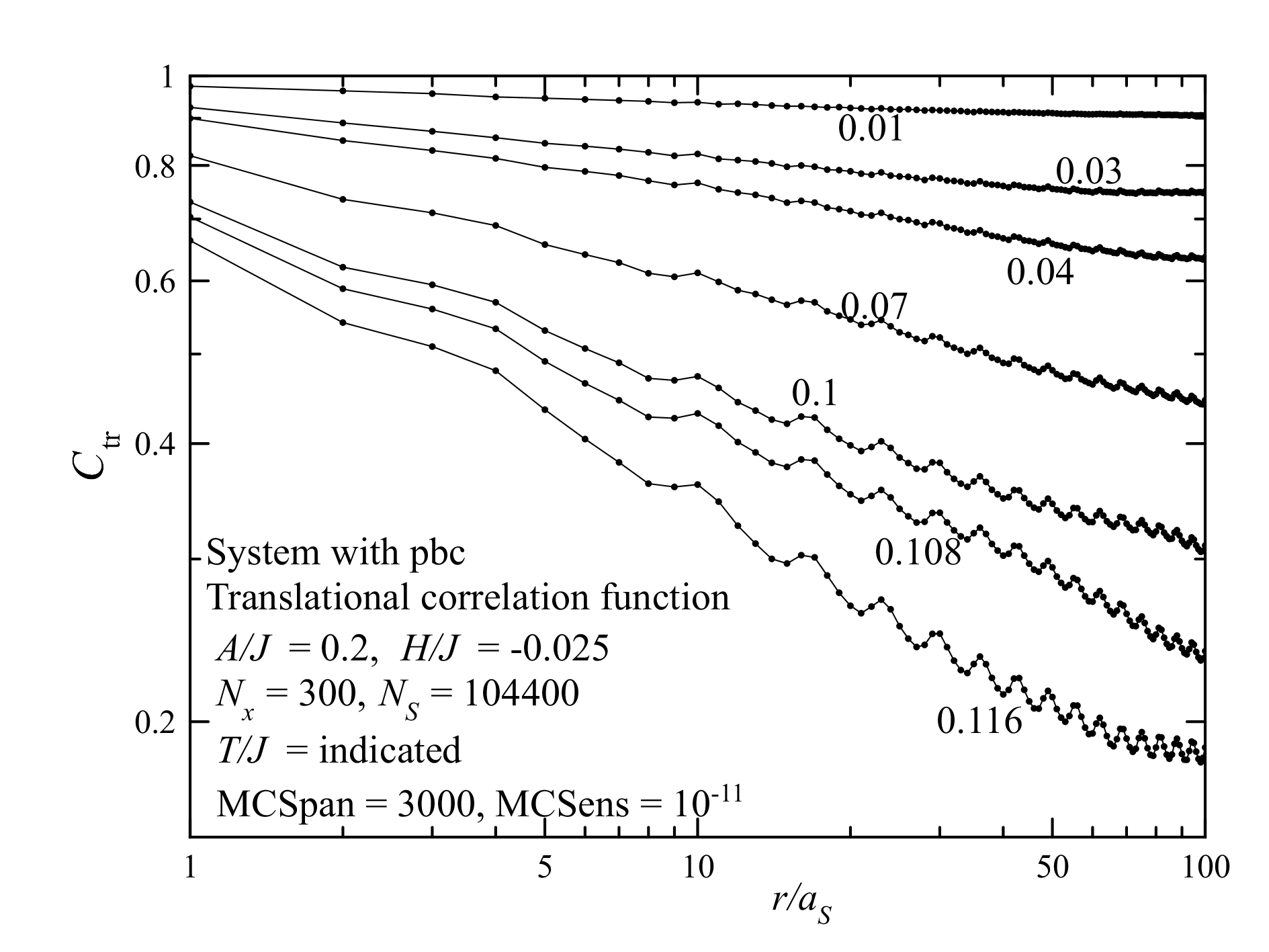}
\caption{Translational correlation function in the skyrmion lattice at different
temperatures.}
\label{Fig-CF_tr}
\end{figure}

\begin{figure}[ht]
\begin{centering}
\includegraphics[width=9cm]{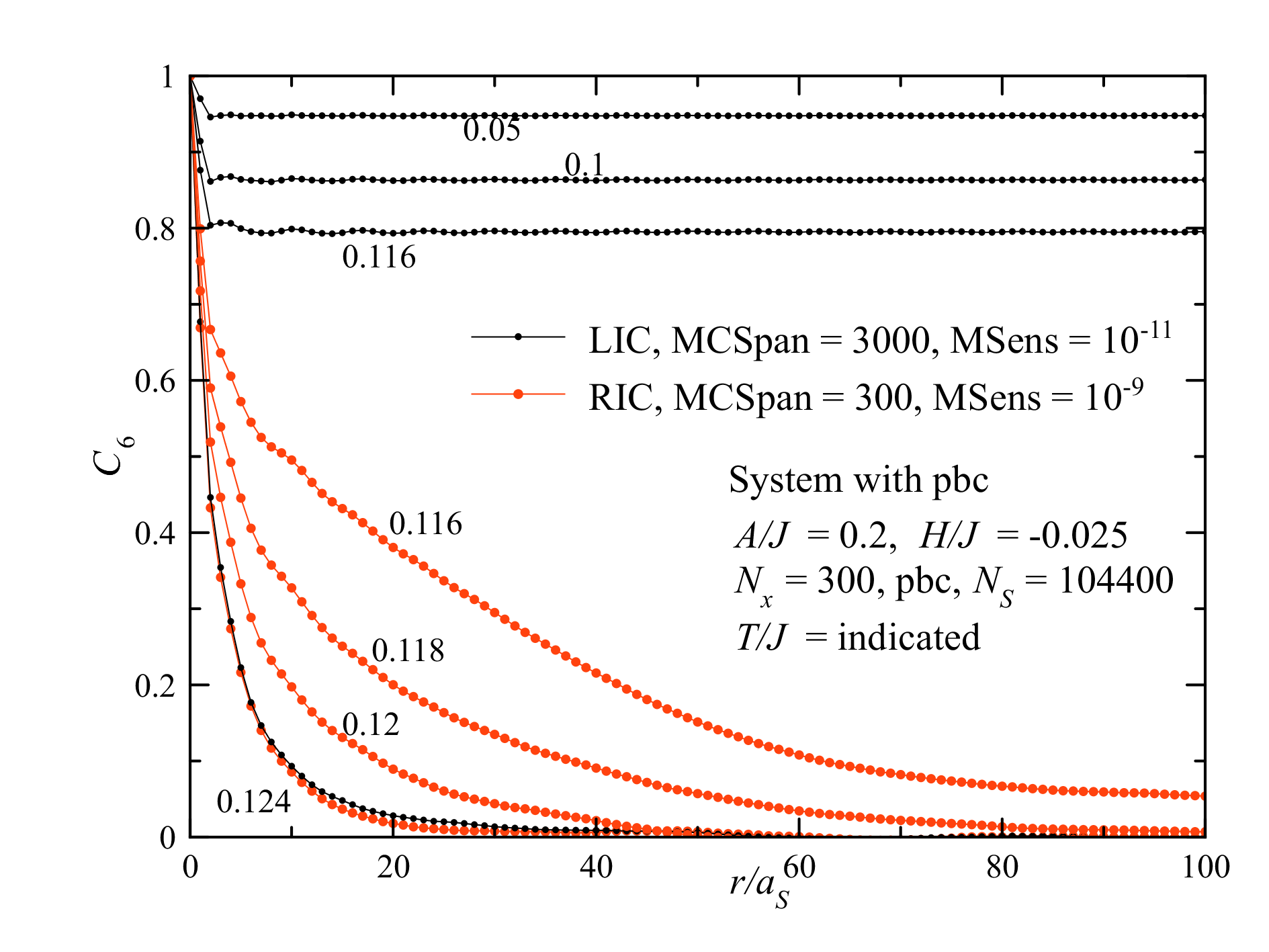}
\par\end{centering}
\begin{centering}
\includegraphics[width=9cm]{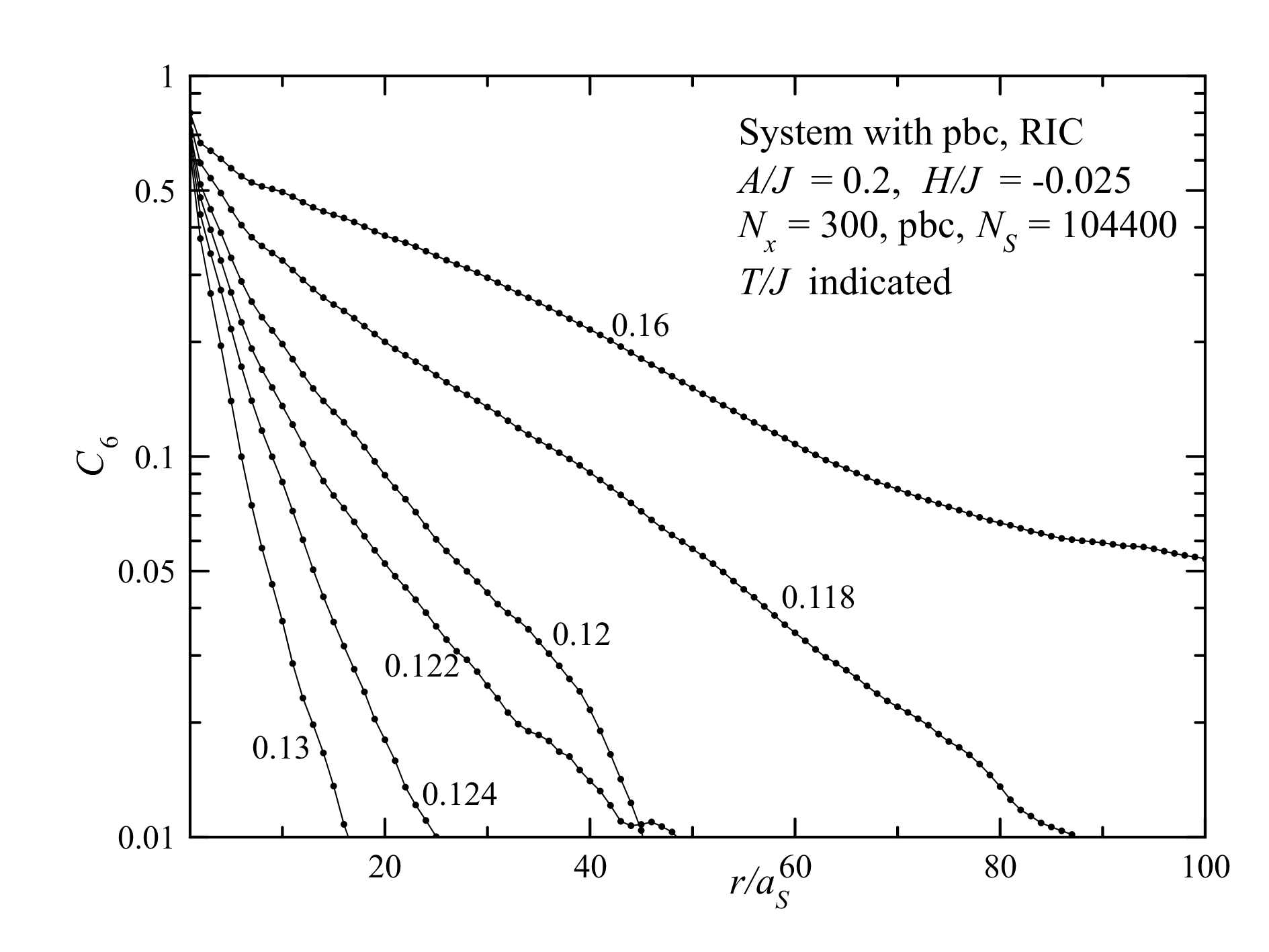}
\par\end{centering}
\centering{}\caption{Orientational correlation function in the skyrmion lattice at different
temperatures. Upper panel: the results for LIC and RIC in the linear
scale. Lower panel: the RIC results in the lin-log scale.}
\label{Fig-CF_hex}
\end{figure}
\begin{figure}
\begin{centering}
\includegraphics[width=9cm]{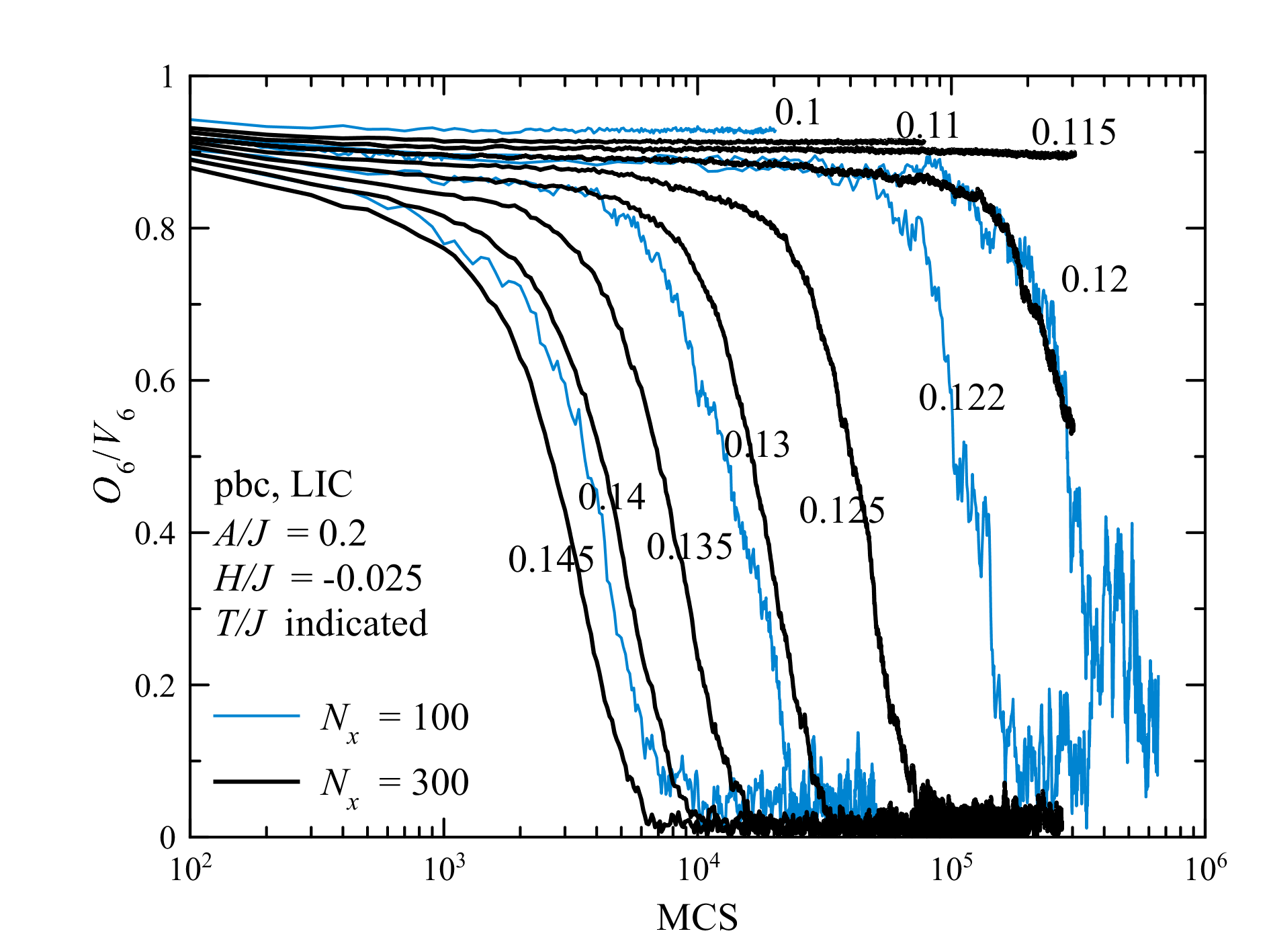}
\par\end{centering}
\caption{Evolution of the orientational order parameter $O_{6}$ in the course
of melting at different temperatures. }

\label{Fig-O6_vs_MCS_pbc_LIC_log-lin}
\end{figure}
\begin{figure}
\begin{centering}
\includegraphics[width=9cm]{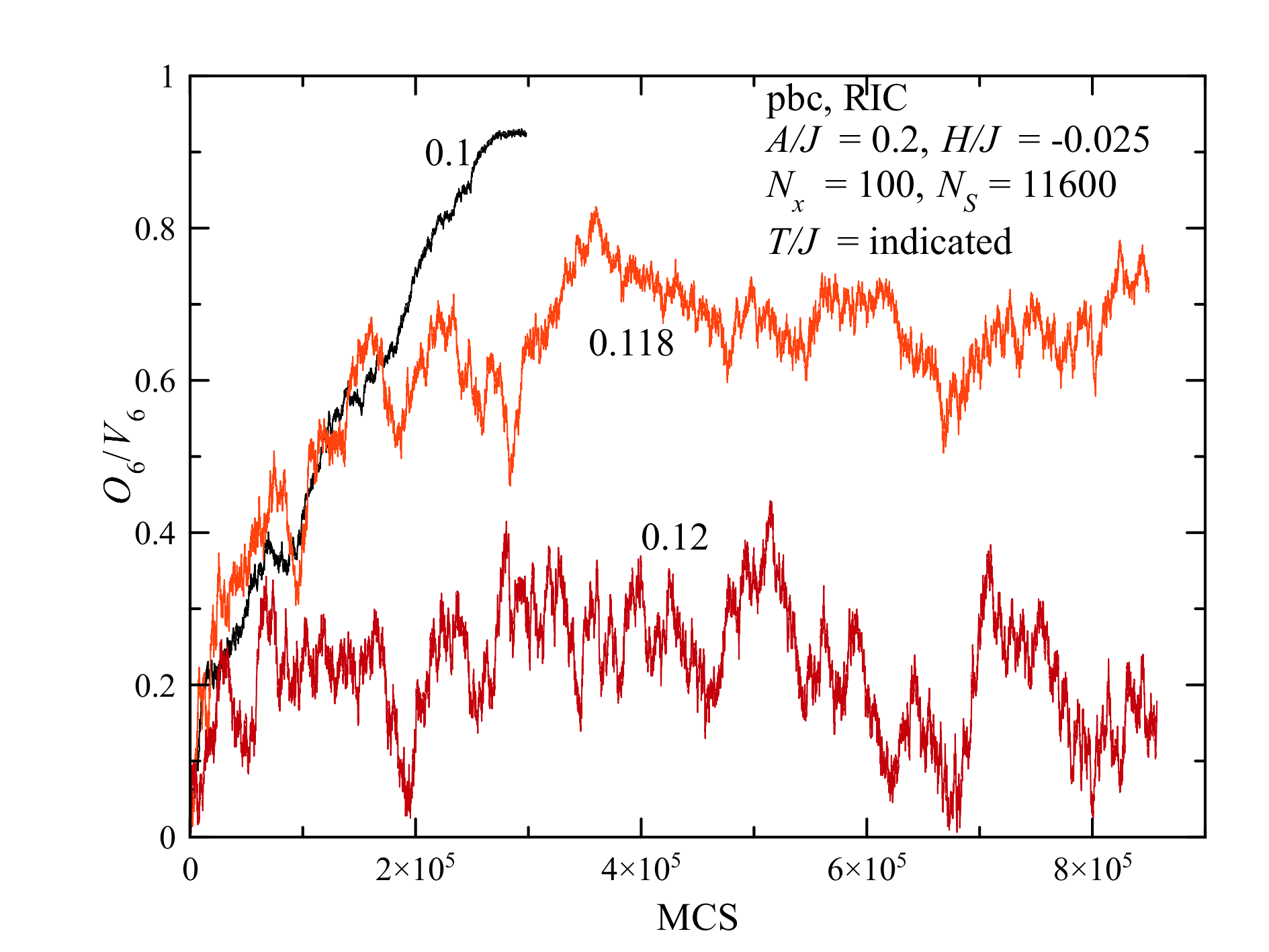}
\par\end{centering}
\caption{Stationary fluctuations of the orientational order in the polyhexatic
phase near the melting transition.}

\label{Fig-Polyhexatic_fluctuations}
\end{figure}
\begin{figure}
\begin{centering}
\includegraphics[width=9cm]{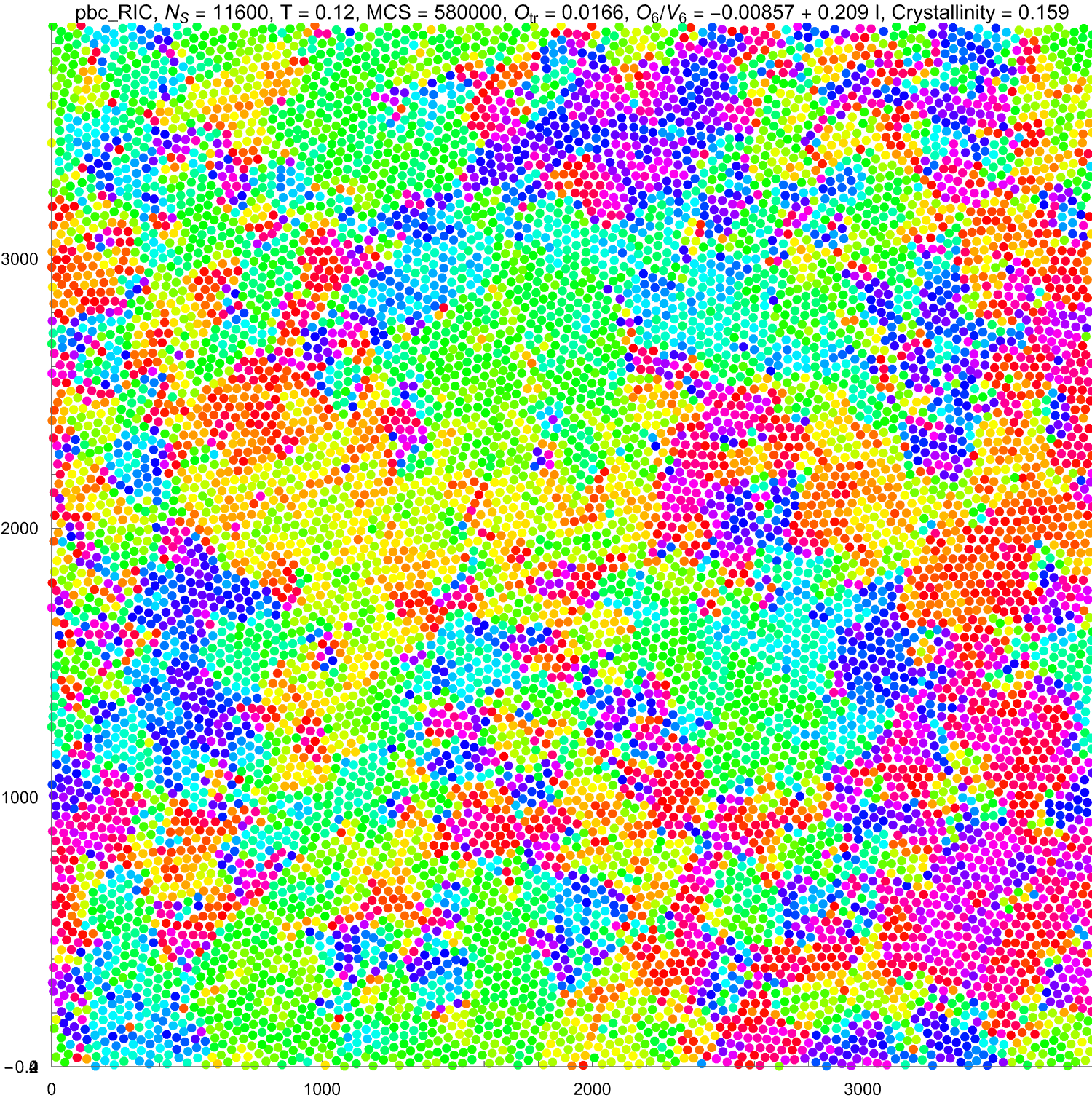}
\par\end{centering}
\begin{centering}
\includegraphics[width=9cm]{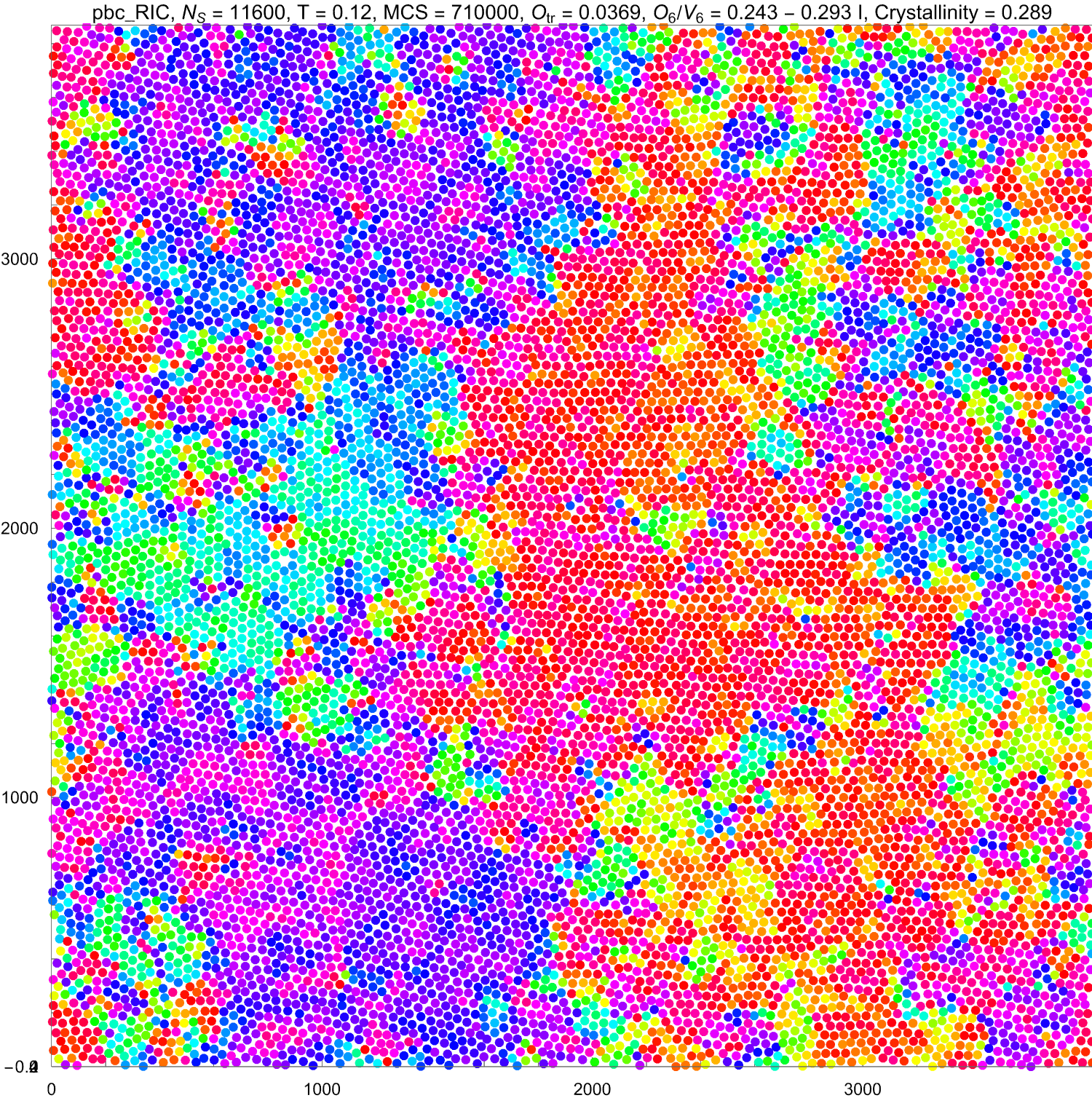}
\par\end{centering}
\caption{Polyhexatic state at $T/J=0.12$ obtained from RIC in the course of
the MC simulation (see Fig. \ref{Fig-Polyhexatic_fluctuations}).
Upper panel: 580000 MCS; Lower panel: 710000 MCS. See the video of
the evolution between these states in the Supplemental Materials.}

\label{Fig-polyhexatic_structure}
\end{figure}

\subsection{Correlation functions}

\label{subsec:Correlation-functions}

The translational correlation function $C_{\mathrm{tr}}(r)$ defined
by Eq. (\ref{Ctr_def}) makes sense only in the LIC case where the
reciprocal-lattice vectors are defined by the perfect lattice used
as the initial state {[}see Eq. (\ref{Reciprocal-lattice_vectors}){]}.
Cooling the system from the melt leads to a polycrystalline state
for which there is no procedure of finding $C_{\mathrm{tr}}(r)$.
Formally computing $C_{\mathrm{tr}}(r)$ with the reciprocal-lattice
vectors mentioned above leads to $C_{\mathrm{tr}}(r)=0$ everywhere.
From the theory of harmonic solids in 2D follows that there is no
translational long-range order because it is destroyed by long-wavelength
fluctuations. At low temperatures the translational CF decreases as
a power of the distance. There is a fair agreement of the numerical
data shown in Fig. \ref{Fig-CF_tr} with the theoretical predictions
-- the behavior of $C_{\mathrm{tr}}(r)$ is approximately a power
law at least up to the distance of a hundred of the skyrmion-lattice
periods $a_{S}$. Note that these fair-quality results were obtained
with very demanding Monte-Carlo parameters $\mathrm{MCSpan}=3000$
and $\mathrm{MCSense}=10^{-11}$ that results in a large number MCS
shown in Fig. \ref{Fig-MCS_vs_T_pbc_Nx=00003D300} (about $5\times10^{5}$
MCS).

Similar results were obtained in many publications on 2D systems of
particles of much smaller sizes, such as $10^{4}$ particles. Our
simulations on such small systems yield rather noisy results, especially
for CFs. To obtain better results for CFs, one should increase the
system size. However, it is difficult to increase the range of distances
in which the power law is seen. At long distances correlations are
establishing very slowly. One needs to perform an exceedingly large
number of MCS to make large portions of the lattice move in different
directions to destroy the long-range order and establish power-law
correlations. In our simulations, $C_{\mathrm{tr}}(r)$ goes to a
plateau at large distances that is an artifact of using a perfect
lattice as the initial state. One also can do simulations with pre-heating
a lattice up to a higher temperature below the melting point, so that
the lattice is not destroyed but translational correlations are strongly
suppressed. Then, thermalization at a lower temperature yields a translational
CF similar to that obtained with LIC at shorter distances but having
random values at larger distances. These considerations show that
it is difficult to reproduce the size dependence of the translational
quasi-order-parameter, Eq. (\ref{Otr_vs_L}), because for this one
needs to know the translational CF at large distances.

The results for the orientational correlation function $C_{6}(r)$,
defined by Eq. (\ref{C6_def}), are shown in Fig. \ref{Fig-CF_hex}
for the LIC and RIC cases. As below the melting point there is a robust
orientational long-range order, $C_{6}(r)$ in the LIC case goes to
a temperature-dependent plateau. Above the melting point in the LIC
case, the behavior of $C_{6}(r)$ changes abruptly and $C_{6}(r)$
goes to zero very quickly. At $T/J=0.124$ there is the same exponential
CF in both LIC and RIC cases. Starting from RIC at temperatures below
melting, one ends up in a polycrystalline state where $C_{6}(r)$
has a larger range characterizing the size of crystallites. The form
of $C_{6}(r)$ here is close to exponential, as can be seen in the
lower panel of Fig. \ref{Fig-CF_hex}.

\subsection{Evolution of the skyrmion lattice during melting and solidification}

In this section we present the time-resolved data on melting and solidification.
The role of time is played by the number of Monte Carlo steps (MCS).
Both for melting and solidification, there are two stages of the evolution.
In the first very fast stage (from 10 to 100 MCS) a partial disordering
or partial ordering occurs in the LIC or RIC cases, respectively.
These changes are due to the rearrangement of skyrmions at shortest
positions. 

Then in the RIC case, small grains of the same hexagon orientation
emerged as the result of the first stage slowly grow, and larger grains
consume smaller ones. This is a very slow process of coarsening via
the motion of grain boundaries. Forming a single crystal for large
systems takes an exceedingly long time, and in reality this process
will be stopped by pinning. The evolution of physical quantities,
such as $0_{6}$, in medium-size systems of about $10^{5}$ skyrmions
is very irregular as the system is split into only a handful of grains
and statistical scatter is very high. To study the coarsening process,
a larger system size is needed.

The second stage of melting apparently includes thermal activation.
Small grains of different hexagon orientation emerge here and there,
then some of them disappear again and some of them grow. Grain boundaries
are moving back and forth. Overall, the initially present single crystal
breaks up in many smaller crystallites. Evolution of the orientational
order parameter in the course of melting at the second stage at different
temperatures are shown in Fig. \ref{Fig-O6_vs_MCS_pbc_LIC_log-lin}.
Closer to the melting point, such as at $T/J=0.12$, melting becomes
a very slow process, so that no substantial melting occurs before
$10^{5}$ MCS.

\subsection{The nature of melting and polyhexatic state of the skyrmion lattice}

The results of our simulations show that melting of the skyrmion lattice
occurs via a first-order phase transition of breaking the lattice
into plates of solid, or grains or domains, with different orientations
of hexagons. Small grains of different orientations emerge as the
result of thermal agitation everywhere in the lattice here and there
but at the temperatures below the melting point $T_{m}$ they disappear
again as there is an apparent free-energy barrier to overcome for
nucleating a surviving grain. The value of $T_{m}$ we obtain slightly
decreases on the simulation length, and its best value is $T/J=0.118$,
obtained for the system with pbc. As melting is extremely slow, it
is difficult to obtain by simulations the true value of $T_{m}$ at
which the free energies of the two phases become equal to each other
and melting time becomes infinite.

The phase above the melting point is not yet a liquid phase because
of a significant size of grains of solid. With increasing the temperature,
these grains become smaller and the system approaches the liquid state.
With lowering the temperature, the grains grow in size. This coarsening
process is very slow and does not lead to a single crystal for large
systems or for systems with frustrating boundaries. The resulting
state at low temperatures is polycrystalline, as shown on Fig. \ref{Fig-Circular_system}.

In the temperature region near the melting point, the state with large
grains of solid is thermodynamically stable and shows eqiulibrium
fluctuations in which boundaries between the grains are moving back
and forth (see a video of the evolution of the polyhexatic state at
$T/J=0.12$ in the course of MC simulation in Supplemental Materials).
With time, the shape of the grains changes significantly but, on average,
they do not become larger or smaller, see Fig. \ref{Fig-Polyhexatic_fluctuations}.
We call this state \textit{polyhexatic state}. The polyhexatic phase
in the system with pbc near melting is shown in Fig. \ref{Fig-polyhexatic_structure}
at two different moments of the MC simulation. The video of the evolution
between these two states can be found in the Supplemental Materials.

\section{Conclusions}

\label{Sec-Conclusion}

In this paper, a simplified approach to the skyrmion lattice is proposed
that is based on the skyrmion's core energy and the recently established
exponential repulsion of skyrmions \citep{CGC-JPCM2020}. With these
two components, one can find the equilibrium concentration of skyrmions,
that is, the period of the skyrmion lattice $a_{S}$. After that,
the number of skyrmions can be fixed as the processes of skyrmion
creation/annihilation are rare at low temperatures where the melting
of the skyrmion lattice takes place. Thus, one can consider skyrmions
as interacting particles and use Monte Carlo or other methods to describe
their properties. This approach is much simpler than considering the
system in terms of spins on the lattice \citep{Nishikawa-PRB2019}.

The results obtained point to a first-order transition of breaking
a single-crystal lattice into the polyhexatic state of large grains
with different orientations of hexagons. With increasing/decreasing
the temperature these grains shrink/grow. The polyhexatic state differs
from the polycrystalline state at $T=0$ by the equilibrium thermal
fluctuations that change its shape.

We haven't observed two transitions with an intermediate state with
a power-law dependence of the orientational correlation function.
In the single-crystal state below melting, this CF has a plateau while
in the polyhexatic state its decay with the distance is closer to
exponential. The CF data near melting are not that accurate because
the ratio of the grain size to the system size is not small enough.
Better results could be obtained with a larger system size. In numerical
work, the direct checking of the KTHNY scenario of the unbinding of
dislocation pairs, followed by the dissociation of disinclination
pairs, is hampered by the difficulty to identify dislocations and
disclinations because at elevated temperatures thermal fluctuations
create significant non-topological local distortions of the crystal
lattice. 

Over the period of forty years, there has been a significant controversy
over the KTHNY scenario in real and numerical experiments, including
the most recent work on skyrmion lattices. Some of it reflects the
differences in the interactions of particles forming the crystal and
the differences in the crystal symmetry {[}53-57{]}. Our results demonstrate
the importance of another factor that is always present in real and
numerical experiments on melting and solidification - nonequlibrium
effects. Real solids rarely form monocrystals. This is true in both
3D and 2D. When the numerical experiment is performed on a large system,
and it resembles real experiment in a sense that the solid is formed
by freezing the liquid, the most common practical result is a polycrystalline
state in which global orientational order is lost. Similarly, when
a sufficiently large 2D crystal melts, there is no guarantee that
the resulting state would be homogeneous with, e.g, power-law decay
of orientational correlations. As our work demonstrates, it consists
of differently oriented domains of the Halperin-Nelson hexatic phase,
which we call a polyhexatic state. Same as for a polycrystalline state,
the orientational order in the polyhexatic state is limited to the
size of the domain. The average size of the fluctuating hexatic domain
goes down on raising temperature above the melting transition, leading
to a fully disordered liquid state at high temperature. 

Quenched disorder is known to have a stronger effect on long-range
correlations than thermal fluctuations. It has been shown that for
a 2D atomic layer on a substrate, the defects in the substrate can
generate a hexatic state characterized by short-range translational
order and algebraic decay of the orientational order even at zero
temperature \citep{EC-PRB1986,Murray-PRL1990,EC-RD-PRB1998}. Thus,
monocrystallinity of the skyrmion lattice may not be sufficient in
search for the temperature-induced hexatic phase. The perfection of
the underlying atomic lattice must be pursued as well.

We hope that further experiments on skyrmion lattices will shed light
on the existence of polycrystalline and polyhexatic 2D phases investigated
in this paper.

\section{Acknowledgments}

This work has been supported by the grant No. DE-FG02-93ER45487 funded
by the U.S. Department of Energy, Office of Science.

\end{document}